\documentstyle[aps,prb,preprint,tighten,eqsecnum,epsfig]{revtex}
\begin{document}

\def\bea{\begin{eqnarray}}
\def\eea{\end{eqnarray}}

\title{\bf Critical behaviour of a spin-tube model in a magnetic field}

\author{R. Citro$^{\star}$, E. Orignac and N. Andrei}

\address{{\it Serin Laboratory, Rutgers University, P.O. Box 849,
Piscataway, NJ 08855-0849, USA}}

\author{ C. Itoi$^{\star \star}$ and S. Qin$^{\star \star \star}$}

\address{{\it Department of Physics and Astronomy, University of British
Columbia\\
 Vancouver, BC, V6T1Z1, Canada}}

\date{\today}

\maketitle

\begin{abstract}
We show that the low-energy physics of the spin-tube model in presence
of a critical magnetic field can be described by a broken  SU(3) spin chain.
Using the Lieb-Schultz-Mattis
Theorem we characterize the possible magnetization plateaus and study 
the critical behavior in the region of transition between
the plateaus $\langle S^z \rangle=1/2$ and $\langle S^z \rangle=3/2$ 
by means of renormalization group calculations
performed on the bosonized effective continuum field theory.
We show that in certain regions of the parameter space of the
effective theory the system 
remains gapless, and we compute the spin-spin correlation functions in
these regions. 
We also discuss the possibility of a plateau at $\langle
S^z\rangle=1$, and  show that although there exists  in the
continuum theory a term that might cause the appearance of a plateau there,
such term is unlikely to be relevant. This conjecture is proved by
DMRG techniques. The modifications of the three-leg ladder 
Hamiltonian that
might show plateaus at $\langle S^z \rangle =1,5/6,7/6$ are discussed, and
we give the expected form of correlation functions on  the $\langle S^z
\rangle=1$ plateau. 
\end{abstract}

\vspace*{20mm}

\noindent PACS numbers: 75.10.Jm, 75.30.Kz, 75.40.Gb \newline
\noindent {\it keywords:} SU(3) spin-chain, renormalization
group, magnetization plateaus, Density Matrix Renormalization Group

\newpage

\section{Introduction}

One-dimensional and quasi-one dimensional quantum spin systems have
attracted much attention in recent years due to the large
number of experimental realizations of such systems
and the variety
of theoretical techniques, both analytical and numerical, available to
study the relevant models. Due to the presence of  large quantum
fluctuations in low dimensions, these systems present unusual properties
such as a gap between a singlet ground state and excited
non-singlet states. Examples include  spin ladder systems in
which a small number of one-dimensional spin-1/2 chains interact among
themselves\cite{ladder}. In this case, in a way very similar to the
Haldane spin-S problem\cite{haldane_gap}, it has been found that if the number
of
chains is even the system effectively behaves as an integer spin
chain with a gap in the low-energy spectrum, while it remains
massless for an odd number of chains. Some two-chain ladders
which exhibit a gap are 
$SrCu_2O_3$\cite{azuma_srcuo} and
$Cu_2(C_5H_{12}N_2)_2Cl_4$\cite{chaboussant_cuhpcl}, and an example
of a gapless 
three-chain ladder is $Sr_2Cu_3O_5$\cite{azuma_srcuo}. Thus far we
implicitly assumed that the boundary conditions in the transverse
direction were open boundary conditions (OBC). These boundary
conditions correspond to having all the chains lying in the same
plane. This is the situation encountered in experimental systems such
as $Sr_2Cu_3O_5$. 
In contrast with OBC, periodic boundary
conditions (PBC) are frustrating for (2n+1) coupled chains. 
As a consequence all the spin
excitations are gapped\cite{schulz_berlin,orignac_spintube} in the
case of periodic boundary conditions. They are also
gapped for 2n chains with PBC but the mechanism is related to singlet
formation as in the OBC case and not frustration. The PBC could be
achieved in an experimental system by having the coupled chains
forming a cylinder instead of lying in a plane.

A richer behaviour emerges when these gapped
or ungapped systems are placed in a magnetic field. Then it is possible for an
integer spin chain to be gapless and a half-odd-integer spin chain to
show a gap above the ground state for certain values of the
field\cite{oshikawa,cabra,cabra_unp,cithra}.
This has been demonstrated by several
methods such as bosonization \cite{schulz_berlin,affleck_houches}, perturbation
theory\cite{reigrotzki} or density-matrix renormalization group method
(DMRG)\cite{white,dmrg0,kawan1}. In particular, it has been shown that
spin-1/2 chains and ladders with a gap
undergo a continuous phase transition from a
commensurate zero uniform magnetization phase to an incommensurate
phase with non-zero magnetization\cite{cithra}, and
the magnetization of the system can exhibit plateaus at certain
non-zero values of the magnetic field\cite{cabra,mila_field}. Further,
a striking property of the quantum spin-chains in a uniform magnetic
field pointing along the direction of the axial symmetry
($z$-direction), is the topological quantization of the magnetization
under a changing of the magnetic field\cite{oshikawa}. It was shown
that starting from
a generalized Lieb-Schultz-Mattis (LSM) theorem\cite{lsm}, 
 that translationally
invariant spin chains in an applied field can be gapful without
breaking translational symmetry, only when the magnetization per spin,
$m$, obeys $S-m=integer$, where $S$ is the maximum possible spin in
each unit cell of the Hamiltonian. Such gapped phases correspond to plateaus at
these quantized values of $m$. In Ref. \onlinecite{sen} the behavior of
the magnetization versus magnetic field has been investigated in
details using DMRG techniques 
for  three coupled spin 1/2 chains with both periodic and open
boundary conditions. Plateaus have been obtained at $m=1/2$ and
$m=3/2$ in agreement with Ref. \onlinecite{oshikawa}. Further, for the case of PBC
a small plateau at $m=0$ was also obtained \cite{cabra_unp,kawan1,sen}. 
Finally, there seems to
exist some weak
evidence for a plateau at $m=1$ for PBC\cite{sen}. Strong coupling Low
Energy Hamiltonians for these two systems were also derived in
Ref.~\onlinecite{sen}. 

In this paper, we  investigate a three-chain ladder with PBC
(spin-tube) in presence of a uniform magnetic field 
by using
bosonization and renormalization group techniques. We are concerned by
the transition region between the magnetization plateaus at $m=1/2$
and $m=3/2$. Our analysis
is based on the low-energy effective Hamiltonian (LEH)  
derived for strong coupling between the rungs\cite{sen}. We identify the
 LEH  as an anisotropic
 SU(3) spin chain with symmetry
breaking terms in a longitudinal magnetic field, and analyze its
 low-energy
physics via bosonization and RG techniques. 
This approach allows us to predict the behavior of the spin-spin
correlation functions in this transition region and the NMR relaxation
rate. This also allows for an investigation of the possibility of a
plateau at $m=1$.

The paper is organized as follows. In Sec.\ref{leh}, we recall the
derivation\cite{sen} of  the LEH
and reduce it to  an anisotropic SU(3) spin chain, while
that in OBC ladder reduces to an anisotropic SU(2) spin chain.  
The difference 
between PBC and OBC models becomes obvious in the language of
effective Hamiltonians. 
 In Sec.\ref{magn} we discuss the generalization of the LSM
theorem\cite{lsm} for SU(3) spin chains. We then review the
analysis of the magnetization process of isotropic SU(3) spin chain, 
and discuss the difficulty\cite{sen} in obtaining the position of the
plateaus from the LEH. 
We show there is no gap at $m=1$
in a simple numerical analysis using density matrix renormalization
group method as well. 
 The bosonized Hamiltonian is derived  in
Sec.\ref{weakcoupling}. In section \ref{RG},  we analyze the
low-energy effective Hamiltonian in a weak-coupling limit by
calculating the one loop renormalization
group (RG) in the marginally perturbed SU(3) Wess-Zumino-Witten
model\cite{prokovskii} and discuss the
renormalization group flow.  At weak coupling, the
 flow is to an invariant surface, leading to
gapless excitations above the ground state with no breaking of the
discrete symmetry. Based on the weak coupling renormalization group
analysis and the usual continuity between weak coupling and strong
coupling in one-dimensional systems, we claim  that the spin-tube is
described by a two component Luttinger liquid at low energy and long
wavelength.
In section\ref{correlations}, we discuss the effect of a variation of
the  magnetic
field in that problem, and show that it does not affect the
two component Luttinger liquid behavior. 
We discuss the  analogy of this two component Luttinger liquid
 with the S2 phase of
the bilinear-biquadratic spin 1 chain\cite{fath_biquadratic}.
Then, having established the equivalence with the
two component Luttinger liquid,   we calculate  the spin
correlation functions in the critical region and  the
temperature dependence of the NMR longitudinal relaxation rate
$T_1^{-1}$. We also present a theoretical description of the plateau
at $m=1$ in the framework of bosonization. Comparing this description
with the numerical results of Ref. \onlinecite{sen} we conclude that
the presence of a plateau at $m=1$ is unlikely in the spin-tube. We verify
 our results on the absence of plateaus at $m=1$ using DMRG, and
indicate the modifications of the spin-tube Hamiltonian that could
lead to a plateau.   
Sec.\ref{conclusions} contains the 
concluding remarks.
 Technical details can be found in the
appendices.

\section{Low-energy effective Hamiltonian of the spin-tube}\label{leh}

\begin{figure}
\centerline{\epsfig{file=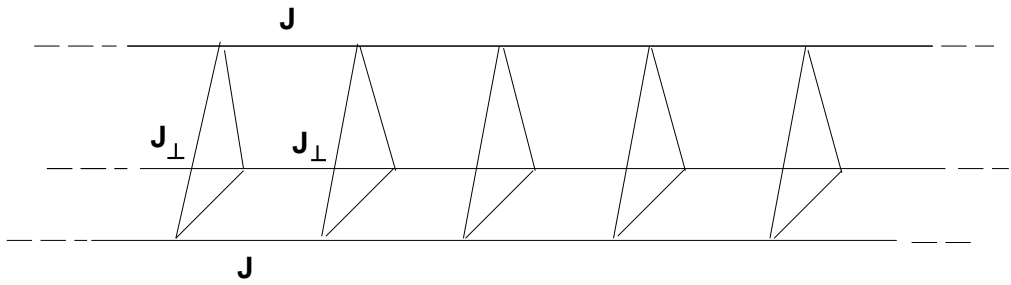,angle=0}}
\caption{Cylindrical three-leg ladder (spin-tube). The choice of the
topology affects the strong-coupling limit.}
\label{fig:tube}
\end{figure}

The Hamiltonian of the three-chain ladder with periodic boundary
conditions (PBC) in the presence of an external
magnetic field is,

\begin{equation}
\label{ham}
H=J\sum_{i=1}^{N}\sum_{p=1}^{3} \vec S_{i,p} \vec S_{i+1,p}+ J_{\perp}
\sum_{i=1}^{N}\sum_{p=1}^{3} \vec S_{i,p} \vec S_{i,p+1} -\vec h
\sum_{i=1}^{N}\sum_{p=1}^{3} \vec S_{i,p},
\end{equation}

\noindent where $p$ (resp. $i$) is a chain (resp. site) index, $J$ is the
coupling along the chain,  $J_{\perp}$ the transverse coupling and
 the site $(i,4)$ is identified with the site $(i,1)$. The three-chain
ladder with periodic boundary conditions can be viewed as forming a tube 
with an equilateral triangular cross section  (see Fig.~\ref{fig:tube}).
We will refer to this system as a {\it spin-tube}.

In the rest of the paper we shall consider the model for
$J_{\perp} \gg J$, and the aim of this section is to recall briefly
 the derivation of the low energy Hamiltonian \cite{sen} in this limit.
To begin with, for $J=0$, the system consists of independent
rungs. The eight states of a given rung fall into a spin-3/2
quadruplet 
and two spin-1/2 doublets. In the absence of a magnetic field, the
spin 3/2 states on a given triangle 
are all degenerate with energy $3/4 J_{\perp}$. These states are:

\begin{eqnarray}
|3/2;3/2\rangle & = & |\uparrow \uparrow \uparrow\rangle \nonumber \\
|3/2;1/2\rangle & = & \frac{1}{\sqrt 3}\lbrack |\uparrow \uparrow \downarrow\rangle +
|\uparrow \downarrow \uparrow\rangle +| \downarrow \uparrow \uparrow\rangle
\rbrack \nonumber \\
|3/2;-1/2\rangle & = &  \frac{1}{\sqrt 3}\lbrack |\downarrow \downarrow \uparrow\rangle+
|\downarrow \uparrow \downarrow\rangle +| \uparrow \downarrow \downarrow\rangle \rbrack
  \nonumber \\
|3/2;-3/2\rangle & = & |\downarrow \downarrow \downarrow\rangle
\end{eqnarray}

Also, in the absence of a magnetic field and on a given rung,  the
two spin 1/2 doublets, corresponding to the left and right
chiralities
(-/+), are degenerate with energy
$-3/4 J_{\perp}$. These states are:

\begin{eqnarray}
|\uparrow +\rangle & = & \frac{1}{\sqrt 3} \lbrack |\downarrow \uparrow
\uparrow\rangle + j |\uparrow \downarrow \uparrow\rangle + j^2 |\uparrow \uparrow
\downarrow\rangle \rbrack \nonumber \\
|\downarrow +\rangle & = & \frac{1}{\sqrt 3} \lbrack |\uparrow \downarrow
\downarrow\rangle + j |\downarrow \uparrow \downarrow\rangle + j^2 |\downarrow \downarrow
\uparrow\rangle \rbrack \nonumber \\
|\uparrow -\rangle & = & \frac{1}{\sqrt 3} \lbrack |\downarrow \uparrow
\uparrow\rangle + j^2 |\uparrow \downarrow \uparrow\rangle + j |\uparrow \uparrow
\downarrow\rangle \rbrack \nonumber \\
\label{doublet}
|\downarrow -\rangle & = & \frac{1}{\sqrt 3} \lbrack |\uparrow \downarrow
\downarrow\rangle + j^2 |\downarrow \uparrow \downarrow\rangle + j |\downarrow \downarrow
\uparrow\rangle \rbrack
\label{doublets}
\end{eqnarray}

\noindent where $j=\exp(\frac{2\pi i}{3})$.

When an external magnetic field is switched on the degeneracy in the different multiplets is lifted.
The energy levels of the
 state $|\uparrow \uparrow \uparrow\rangle$ (in the spin-3/2 multiplet) 
and the spin-1/2 states
$|\uparrow +\rangle$, $|\uparrow -\rangle$ cross at $h_c=\frac{3}{2}
J_\perp$ (see. Fig.~\ref{fig:levels}). As a result for
$h<h_c$ one has a ground state 
magnetization $\langle  S^z \rangle=1/2$ and for $h>h_c$,  $\langle  S^z
\rangle=3/2$, i.e. $h_c$ is a transition point between two
magnetization plateaus. If a  small coupling $J$ is
turned on,  this transition
is expected to broaden between $h_{1/2,+}$ and $h_{3/2,-}$, where
$h_{3/2,-}-h_{1/2,+}$ is of the order of $J$. We expect that in this
interval $\langle  S^z\rangle$ increases continuously with $h$.
 In this limit the properties of
the system can be studied by perturbing with $H_1$ around the decoupled rung
hamiltonian $H_0$,
\bea
H&=&H_0+H_1 \\
H_0&=&J_{\perp}\sum_{i=1}^{N}\sum_{p=1}^{3} \vec S_{i,p} \vec S_{i,p+1}
-h_c \sum_{i=1}^{N}\sum_{p=1}^{3}S_{i,p}^{z}\label{h0},\\
H_1&=& J\sum_{i=1}^{N}\sum_{p=1}^{3} \vec S_{i,p} \vec S_{i+1,p}
-(h-h_c) \sum_{i=1}^{N}\sum_{p=1}^{3}S_{i,p}^{z}.\label{h1}
\eea

 At $h=h_c$ the groundstate of $H_0$ is $3^{N}$ fold degenerate,
 the states  $|\uparrow -\rangle_i$,
$|\uparrow +\rangle_i$, $|\uparrow \uparrow \uparrow\rangle_i$, to be
denoted respectively  $|\tilde 1\rangle_i$, $|\tilde 2\rangle_i$,
$|\tilde 3\rangle_i$, spanning the low-energy subspace. $H_1$
 lifts the degeneracy in the subspace, leading to an
effective Hamiltonian that can be derived by standard perturbation
theory. Since in the truncated subspace there are  3
states per triangle, it is natural  to express the spin
operators in the basis given by  Gell-Mann matrices  $\lambda^{\alpha},\; \alpha= 1 \dots 8$. 
 (The conventions we use for the matrices can be found for instance in
Refs. \onlinecite{gellmann_matrix} and \onlinecite{gasiorowicz}).   
By considering the action of the spin operators $S_{1,2,3}^+$ and
$S_{1,2,3}^z$ on each state  of  the truncated Hilbert space
the spin operators can be expressed in
terms of the  matrices as,
\bea   
S_{i,p}^{+} & =& \frac{1}{2\sqrt{3}}\lbrack j^{p-1} (\lambda_i^{6}+i
\lambda_i^{7})+
j^{2(p-1)}(\lambda_{i}^{4}- i \lambda_{i}^{5})\rbrack \\
S_{i,p}^{z} &= &\frac{1}{3} \lbrack
\frac 5 6 I -\frac{\lambda_i^{8}}{\sqrt{3}}-
j^{2(p-1)}(\lambda_{i}^{1}+i\lambda_i^{2})-j^{(p-1)}(\lambda_{i}^{1}-i\lambda_i^{2}) \rbrack \label{si}
\eea
\noindent where $I$ is the identity matrix. 
The total rung spin is given by: 
\bea\label{siz}
S_i^z=\left ( \frac{5}{6}I -\frac{\lambda_i^8}{\sqrt{3}}\right ).
\eea

The effective Hamiltonian to first order then becomes:
\bea
H_{\text{eff.}}&=&\tilde{H}_0 + \tilde{H}_I, \label{heff1} \\
\tilde{H}_{0}& = &\frac{J}{4}
\sum_{i=1}^{N}\sum_{\alpha=1}^8\lambda_{i}^{\alpha}
\lambda_{i+1}^{\alpha} \label{h_0}\\
\tilde{H}_I&=&q\sum_{i=1}^{N}[\lambda_{i}^1\lambda_{i+1}^1+\lambda_{i}^2\lambda_{i+1}^2] + u \sum_{i=1}^{N} \lambda_i^{3} \lambda_{i+1}^3
+u^\prime \sum_{i=1}^N \lambda_{i}^8 \lambda_{i+1}^8
+ \frac{h_{\text{eff.}}}{\sqrt{3}}
\sum_{i=1}^{N} \lambda_i^{8} \label{h_1}
\eea
In our case, $q=-J/12$, $u=-J/4$, $u^\prime=-5J/36$
and $h_{\text{eff.}}=h-h_c-5J/9$; hereafter we choose our units so that
$J=1$. 
The Hamiltonian (\ref{heff1}) is written as an isotropic SU(3) spin
chain $\tilde H_0$ and terms $\tilde H_I$ that break the symmetry.
This form will 
be convenient later on when we study such questions as what regions of
parameter space are gapless and  the behavior of correlations
functions there. 

Another form of the Hamiltonian is convenient when one wishes to study
the plateau structure.
Introduce\cite{sen} a different basis of $SU(3)$
 operators $T^{\pm}_{1}$,
$T^{\pm}_{2}$, $T^{\pm}_{3}$, and $T^z$ defined by:

\begin{eqnarray}
\label{eq:T-lambda}
T_1^{\pm}&=&(\lambda_1 \pm \imath\lambda_2)/2  \\
T_2^{\pm}&=&(\lambda_4 \pm \imath \lambda_5)/2 \\
T_3^{\pm}&=&(\lambda_6  \pm \imath \lambda_7)/2 \\
T^{z}&=& -2\frac{\lambda_8}{\sqrt{3}}
\end{eqnarray}
                                                              
Then, to first order, and up
to a constant, the effective Hamiltonian  reads:
\begin{eqnarray} \label{heff}
H_{eff} & = & \frac{J}{2} \sum_{i} \lbrack T^{+}_{i,2}T^{-}_{i+1,2}+
T^{-}_{i,2}T^{+}_{i+1,2}+ T^{+}_{i,3}T^{-}_{i+1,3}+
T^{-}_{i,3}T^{+}_{i+1,3}\rbrack+ \nonumber \\
& + & \frac{J}{3} \sum_i \lbrack T^{+}_{i,1}T^{-}_{i+1,1}+
T^{-}_{i,1}T^{+}_{i+1,1} \rbrack +\frac{J}{12} \sum_i
T^{z}_{i}T^{z}_{i+1} -
(\frac{1}{2}h - \frac{3}{4} J_{\perp} -\frac{5}{18} J) \sum_iT^{z}_i.
\end{eqnarray}

This is  the Hamiltonian derived in Ref. \onlinecite{sen}.
 In this form the underlying structure of 
  an anisotropic SU(3) spin chain in a 
``$\lambda^8$ magnetic field''   is unexploited.
 The  correspondance between our notations and those of
 Ref. \onlinecite{sen} can be found in table
 \ref{correspondance1}.

 The form of the Hamiltonian (\ref{heff1})
 may help in relating
our model to integrable versions of the SU(3) spin chains. Isotropic
 spin chains are known to be
integrable by Bethe Ansatz techniques\cite{uimin,sutherland}. The
magnetization process of SU(3)
spin chains with a magnetic field coupled to $\lambda^3$ has been
analyzed in the context of the bilinear-biquadratic spin 1 chain at
the integrable Uimin-Lai-Sutherland point
\cite{parkinson_uls_magfield,kiwata_uls_magfield,schmitt_uls_phase_diagram}
by solving numerically the Bethe Ansatz equations.
There exist also integrable  anisotropic SU(3) spin-chains
\cite{maassarani_su(n)_xx}, but the chain described by the
Hamiltonian (\ref{heff1}) is not one of them.
To study the magnetization process and the correlation functions of
the chain we will therefore have to resort
to a combination of approximate methods such as bosonization,
renormalization group and strong coupling analysis.
We will, first, analyze the strong-coupling
effective Hamiltonian (\ref{heff}) to study the region of transitions
between plateaus of the magnetization, next we will discuss the
low-energy properties of the effective Hamiltonian (\ref{heff1}) via a
renormalization group analysis.

We conclude this section by contrasting the Open and Periodic
boundary conditions. The same strong coupling 
analysis can be done for the OBC case. In contrast with the PBC 
case, we have only a two fold degeneracy instead of three
at $J=0$ under a strong
field $h= \frac{7J_{\perp}}{8}$. 
These two low energy states are
\begin{eqnarray}
|3/2; 3/2 \rangle &=& |\uparrow \uparrow \uparrow \rangle \nonumber \\  
|1/2; 1/2 \rangle &=&\sqrt{\frac{2}{3}} \left(
\frac{1}{2} |\uparrow \uparrow \downarrow \rangle +
\frac{1}{2} |\downarrow \uparrow \uparrow \rangle-
|\uparrow \downarrow \uparrow \rangle.
\right)
\end{eqnarray}
The effective Hamiltonian to first order perturbation in $J/J_{\perp}$
becomes the  well known spin-1/2 XXZ model,
\begin{equation}
H_{eff} = \frac{J}{4} \sum_i \left( \sigma ^x_i \sigma ^x_{i+1} + 
\sigma^y_i \sigma^y_{i+1}
+ \Delta \sigma^z_i \sigma^z_{i+1} \right)-\left( \frac{h}{2}-
\frac{7J_{\perp}}{16} \right) \sum_i \sigma^z_i ,  
\label{obc}
\end{equation}
where $\Delta = \frac{5}{18}$.
It is well known that this Hamiltonian has no gap for $\Delta \leq 1$
except at the saturated magnetization $\sigma^z = \pm 1$,
which corresponds to $m=1/2, 3/2$ in the original ladder model.
This agrees with a weak coupling analysis ($J >> J_{\perp}$)
based on bosonization \cite{schulz_berlin,cabra_unp}.  
In the strong coupling analysis, one can clearly see the  difference
between PBC and OBC in their effective Hamiltonian. 

\section{Analysis of the magnetization process of an anisotropic
SU(3) chain}

We begin the analysis by discussing the generalization of
 the Lieb, Schultz, Mattis
theorem to the SU(3) case\cite{lsm_theorem,affleck_lieb}. The theorem
 allows us to predict
 the possible locations of the plateaus. Comparing then these
predictions for a chain of SU(3) spins for the plateaus' location with
 those
for 3 spin-1/2 chains we conclude that the mapping onto a
SU(3) spin chain does not introduce spurious plateaus, and should
therefore give physically correct results for $|h-h_c|\sim J$.
Next, we will discuss the magnetization process of an isotropic
 SU(3) spin  chain and address the question whether a cusp  appears
 in the magnetization versus magnetic
field curve (such a cusp is not related to magnetization plateaus). 
A cusp has been observed in spin-1 chains with
bilinear-biquadratic exchange which can be mapped onto an  SU(3)
spin chain  for a special value of the biquadratic exchange.
We argue that in our case a cusp in the magnetization should not be expected. 
Finally, We recall how the values of the magnetic field corresponding
to the plateaus\cite{sen} at $\langle S^z \rangle=1/2,3/2$ are
obtained from the anisotropic SU(3) spin chain. 

\subsection{ The Lieb-Schultz-Mattis Theorem for a chain of SU(3)
spins}

Consider an anisotropic chain of SU(3) spins,
\begin{equation}\label{eq:generic_anisotropic_su3}
H=\sum_{i=1}^N \left[\sum_\alpha a_\alpha \lambda^\alpha_i
\lambda_{i+1}^\alpha + \beta (\lambda_i^3 \lambda_{i+1}^8+\lambda_i^8
\lambda_{i+1}^3) \right],
\end{equation}
where $a_1=a_2, a_4=a_5, a_6=a_7$, which we rewrite as,
\begin{eqnarray}
H & = &   2 \sum_i \left[ a_1
(T_{i,1}^+ T_{i+1,1}^- +T_{i,1}^- T_{i+1,1}^+)+a_4 (T_{i,2}^+
 T_{i+1,2}^- +T_{i,2}^- T_{i+1,2}^+) +a_6 (T_{i,3}^+ T_{i+1,3}^- +T_{i,3}^- 
T_{i+1,3}^+)+ \right .
\nonumber \\
\label{eq:hamiltonian_tuv}
&  & \;\;\; + \left . 2a_3 \lambda_i^3\lambda_{i+1}^3 + 2a_8 \lambda_i^8
\lambda_{i+1}^8 + 2 \beta (\lambda_i^3 \lambda_{i+1}^8+\lambda_i^8
\lambda_{i+1}^3)\right] .
\end{eqnarray}
The anisotropic SU(3) chain we
are considering in this paper falls into this class of Hamiltonians.

The purpose of the LSM theorem is to classify possible gapless excitations 
above the ground state.
Introduce the operators:
\begin{eqnarray}
U_3 & = & \exp\left(\imath \frac{2\pi}{N} \sum_{n=1}^N n \lambda_n^3\right)
\label{U3} \\
U_8 & = & \exp\left(\imath \frac{2\pi}{N\sqrt{3}}
\sum_{n=1}^N n \lambda_n^8\right)\label{U8}
\end{eqnarray}
\noindent and  begin by  studying the state $U_3| 0 \rangle$. We wish to
compare its energy expectation value with the vacuum's.
Consider therefore the 
expression $\langle 0 | U_3^\dagger H U_3 -H | 0 \rangle$.
Using,
\begin{eqnarray}\label{eq:transformation_u3}
U_3^\dagger T_{n,1}^+ U_3& = &  e^{2\imath
\frac{2\pi n}{N}} T_{n,1}^+ \nonumber \\
U_3^\dagger T_{n,2}^+ U_3 & = &  e^{\imath
\frac{2\pi n}{N}}T_{n,2}^+ \nonumber \\
U_3^\dagger T_{n,3}^+ U_3 & = &  e^{-\imath
\frac{2\pi n}{N}}T_{n,3}^+
\end{eqnarray}
we find upon developing it as a power
series in $1/N$ ($N$ is the size of the system),  that the zeroth 
order term vanishes, since it  contains averages of the form,
$\imath \langle T_{n+1,p}^+ T_{n,p}^- -T_{n+1,p}^-
T_{n,p}^+\rangle$, which cancel when invariance under parity $m \to 2n+1-m$, 
is used: $\langle T_{n+1,p}^+ T_{n,p}^- \rangle = \langle T_{n+1,p}^-
T_{n,p}^+\rangle$. Therefore, the expansion begins with a first
 order term in $1/N$,
\bea
\frac 1 N \times \left(-\frac {2\pi^2}{N} \sum_n \left[ 4 a_1 \langle 
T_{n+1,1}^+
T_{n,1}^- + H. c. \rangle + a_2 \langle T_{n+1,2}^+
T_{n,2}^- + H. c. \rangle + a_3 \langle T_{n+1,3}^+
T_{n,3}^- + H. c. \rangle\right] \right) \nonumber
\eea

Translation invariance guarantees that the coefficient of $1/N$ in
this term is indeed finite. The translation  operator $T$,
 is defined by:
\begin{equation}
T^{-1}\lambda_n^\alpha T=\left \{\begin{array}{cc}
\lambda_{n+1}^\alpha & \mbox{ }\mathrm{if }\; n<N \\
\lambda_1^\alpha & \mbox{ }\mathrm{if }\; n=N, \\
\end{array}\right.
\end{equation}
and translation invariance,  $T |0 \rangle=  |0 \rangle$, (together 
with the expression
of $U_3$,  Eq.(\ref{U3})), implies:
 $T U_3 |0 \rangle= T U_3 T^{-1} T |0 \rangle=e^{-2 \pi \imath
(\lambda_N^3-m_3)}| 0 \rangle$, where
$m_3=\frac 1 N \sum_{n=1}^N \lambda_n^3$.
Therefore, the state $U_3\mid 0 \rangle$ is orthogonal to the ground
state if $m_3$ is non-integer. This state has energy $O(1/N)$ above the
ground state, indicating either a ground state with a broken
symmetry or gapless excitations above the ground state for m
non-integer\cite{lsm,oshikawa}. 
A gap in the excitation spectrum can only exist for $m_3=-1,0,1$ in the
absence of broken symmetry ground states.
For a ground state with $p$-site-periodicity
instead of one site translational symmetry,
\begin{equation}
T^p |0 \rangle = |0 \rangle,
\end{equation}
we obtain,
$T^p U_3 | 0 \rangle = e^{- 2 \pi \imath (\sum_{i=0} ^p 
\lambda^3 _{N-i} - p m_3)}$ for a positive integer $p$. 
A gap excitation can then
appear only for $m_3= q/p$ with an integer $q= -p, \cdots, p$.

If we now consider the action of $U_8$, we have:
\begin{eqnarray}\label{eq:transformation_u8}
U_8^\dagger T_{n,1}^+ U_8 & = &   T_{n,1}^+ \nonumber \\
U_8^\dagger T_{n,2}^+ U_8 & = &  e^{\imath
\frac{2\pi n}{N}}T_{n,2}^+ \nonumber \\
U_8^\dagger T_{n,3}^+ U_8 & = &  e^{\imath
\frac{2\pi n}{N}}T_{n,3}^+
\end{eqnarray}

\noindent Once again,  $\langle  0\mid  U_8^\dagger H U_8 -H \mid
0\rangle = O(1/N)$, but this time: $TU_8\mid 0
\rangle=e^{-\imath \frac{2\pi}{\sqrt{3}}(\lambda_8^N-m_8)}U_8 \mid 0 \rangle$,
where  $m_8=\frac 1 N \sum_{n=1}^N \lambda_n^8$.
This implies that a gapful excitation on a translational
invariant ground state is only possible for
$\frac{m_8}{\sqrt{3}}=n +1/3$, where $n$ is integer.
As in the case of $U_3$, a gap on $p$-site-periodic ground state 
can exist for 
$p \frac{m_8}{\sqrt{3}} =n + 1/3$.
As a consequence, two conditions have to be met to avoid having
gapless excitations above the translationally invariant ground state (p=1):
\begin{eqnarray}\label{eq:conditions_plateaus}
\frac{m_8}{\sqrt{3}}&=&n +1/3 \nonumber \\
m_3&=&n^\prime.\nonumber
\end{eqnarray}
The preceding discussion is quite general. In the specific case we are
considering, the magnetic excitations are described by
$S^z=5/6- \lambda_8/\sqrt{3}$. Therefore, we expect that the
magnetization plateaus are associated with the absence of excited
states above the ground states generated by $U_8$. This implies that
the only possible magnetization plateaus correspond to $\lambda_8$.
Clearly, such a conclusion could have   been reached   by considering
the LSM Theorem for a three chain system. 
However, the LSM theorem for the SU(3) chain also indicates the possibility
of a gap for chiral excitations for $\langle \lambda_3 \rangle =
-1,0,1$. Because the Hamiltonian (\ref{h_0})--(\ref{h_1})
 shows chiral symmetry, one has necessarily $\langle \lambda_3
\rangle=0$. Therefore, according to the LSM theorem, we are in a situation 
where a gap in the chirality modes can be present and we should decide
whether this gap actually opens. Let us consider two simple limiting
cases.
 For $\langle \lambda_8 \rangle/\sqrt{3}=1/3$,
i.e. $\langle S^z \rangle=1/2$, the only possible states on each site
of the SU(3) spin chain are $\mid \tilde 1 \rangle$ and$\mid \tilde 2
\rangle$. In that case, it can be easily seen that the
Hamiltonian (\ref{heff}) reduces to an effective XY chain \cite{sen},
implying gapless chiral excitations. On the other hand, if
$\langle \lambda_8 \rangle/\sqrt{3}=-2/3$, the only remaining mode on
each site is $|\tilde 3 \rangle$. One has indeed in this state
$\lambda_3=0$, but this time there is a gap in chiral excitations.
If there is a ground state with broken translational symmetry,
the situation is more complicated.
Therefore, the question of actual gap opening cannot be settled by the
LSM theorem alone. 

\subsection{Comparison with the magnetization process of an isotropic
SU(3) spin chain}
The isotropic SU(3) spin chain is known to be integrable by the Bethe
Ansatz~\cite{uimin,sutherland}. It is also known that the bilinear
biquadratic spin-1 chain defined by the Hamiltonian:
\begin{equation}\label{eq:bilinear_biquad}
H=\sum_i \left[ \vec{S}_i.\vec{S}_{i+1} +\beta
(\vec{S}_i.\vec{S}_{i+1})^2 -h S^z_i \right]  
\end{equation}
for $\beta=1$ (ULS point) can be mapped onto an isotropic SU(3) spin chain.
In the context of the bilinear-biquadratic spin chain, the
magnetization process of isotropic SU(3) spin chains has been
investigated in
detail by solving
numerically the Bethe Ansatz equations.
In that case, the magnetic field couples to $\lambda_3=S^z$.
A cusp in was obtained in
the magnetization as a function of the magnetic field\cite{parkinson_uls_magfield,kiwata_uls_magfield}. The
 cusp is due to the fact that three different types of Bethe Ansatz
quasiparticles have respectively chemical potentials: $h,0,-h$. As a
result, when $h$ is large enough the band with the highest chemical
potential is emptied causing the cusp in the magnetization.
Also the effect of an
anisotropy $-D(S_i^z)^2 \propto \lambda^8$ on the bilinear biquadratic
spin-1 chain at the ULS point was studied \cite{schmitt_uls_phase_diagram}. 
Rephrasing the results in the context
of SU(3) spin chains, it was found that when one applies a field that
couples to  $\lambda^8$ (respectively $\lambda^3 +\frac
{\lambda^8}{\sqrt{3}}$,  $\lambda^3 -\frac
{\lambda^8}{\sqrt{3}}$)  the resulting average value of $\lambda^8$
 (respectively $\lambda^3 +\frac
{\lambda^8}{\sqrt{3}}$,  $\lambda^3 -\frac
{\lambda^8}{\sqrt{3}}$) shows no cusp as a function of the applied
magnetic field. The reason is that in this case, the Bethe ansatz
particles have chemical potentials $-h,-h,\;2h$ which prevent band
emptying effects.  
 In our case, similarly, the magnetic field couples to
$\lambda_8$. We thus conjecture that although
the anisotropy renders the system non-integrable, the absence of band
emptying effects should persist in the anisotropic case preventing any
cusp in the magnetization. 

Also, on general grounds, the Hamiltonian (\ref{heff}) is invariant
under exchange of chiralities. Therefore, we expect to have $\langle
\lambda_3 \rangle=0$, whatever the applied magnetic field.

\subsection{Strong-coupling analysis of the effective Hamiltonian:
Magnetization plateaus}\label{magn}

In this section we will recall the evolution of the ground state
magnetization as a function of the external magnetic field. As noted
 before, when the intrachain coupling $J$ is set to zero,
upon switching the external magnetic field  on, we find
at increasing $h$ that
the ground state of a given rung undergoes a transition between the
spin-3/2 state, $|\uparrow \uparrow \uparrow\rangle$  and the spin-1/2 states,
$|\uparrow +\rangle$, $|\uparrow -\rangle$ at $h_c=\frac{3}{2} J_\perp$, 
resulting in a sudden jump of the magnetization between $m=1/2$ and
$m=3/2$ (as shown in Fig.~\ref{fig:levels}).
If the coupling $J$ is non-zero but small this transition
is broadened between $h_{3/2,-}$ and $h_{1/2,+}$ which can be identified,
respectively, with the lower and upper critical fields of the saturation
plateaus of the magnetization at $m=3/2$ and $m=1/2$ in the
terminology of ref. \onlinecite{sen}.  As noted there, it is easy to obtain 
\begin{equation}
h_{3/2,-}=(\frac{3}{2} J_{\perp} +2J).
\end{equation}
by considering the condition for the ferromagnetic state $\mid \tilde
3 \tilde 3 \cdots \tilde 3 \rangle$ to be stable upon the
introduction of spin 1/2 states $| \tilde 1 \rangle$ or $|\tilde 2
\rangle$. 

It is harder to determine the lower critical field $h_{1/2,+}$, below which
 the magnetization plateau is at $m=1/2$. Indeed, on the 
plateau $m=1/2$, there are two
possible states on each site, $| \tilde 1 \rangle$ or $|\tilde 2
\rangle$, so that this plateau is described by an
effective XY chain. As a consequence, the ground state wavefunction
for $m=1/2$, obtained from the Jordan-Wigner transformation is a
complicated linear combinations of states of the form $| \cdots
\tilde{2} \tilde{2} \tilde{1} \tilde{2} \tilde{1} \cdots \rangle$.
One has to consider the energy loss created by the introduction of a
$|\tilde 3\rangle$ state in that chain, and balance it with the energy
gained from the magnetic field. This problem bears some
similarity to the dynamics of a few holes
in a $t$-$J$ chain, which has an SU(2)$ \times $U(1) symmetry
instead of the U(1) $\times$ U(1) symmetry in our model.  
 The analogy with the $t$-$J$ model suggests 
a two-component Luttinger liquid behavior of the system in a large
part of the phase diagram.

\section{bosonization and weak-coupling analysis}\label{weakcoupling}

We proceed now to study the long distance properties
of the effective Hamiltonian $H_{\text{eff.}}$ defined in Eq. (\ref{heff1}).
It is a sum of
an isotropic SU(3) spin chain  Hamiltonian  plus SU(3) symmetry
breaking terms,

\bea
\label{htot}
H_{\text{eff.}}=\tilde H_0+\tilde H_1+\tilde H_2+\tilde H_3+\tilde H_h
\eea

\noindent with 

\bea
\tilde H_0&=&\frac{J}{4}
\sum_{i=1}^{N}\sum_{\alpha=1,\dots,8}\lambda_{i}^{\alpha}
\lambda_{i+1}^{\alpha} \label{su3ham} \\
\tilde H_1&=&q\sum_{i=1}^{N}\lbrack \lambda_{i}^1\lambda_{i+1}^1+
\lambda_{i}^2\lambda_{i+1}^2 \rbrack \label{lamb67}\\
\tilde H_2&=&u\sum_{i=1}^{N}\lambda_{i}^3\lambda_{i+1}^3 \label{lamb33} \\
\tilde H_3&=&u'\sum_{i=1}^{N}\lambda_{i}^8\lambda_{i+1}^8 \label{lamb88} \\
\tilde H_h&=&h_{eff}\sum_{i=1}^{N}\frac {\lambda_i^8}{\sqrt{3}} \label{lambh}
\eea

\noindent In our case, $q=-1/12 J$, $u=-1/4 J$,
$u'=-5/36 J$,  and
$h_{eff}=\frac{h-h_c}{2}-\frac{5}{18}J$;
 hereafter we choose our units so that $J=1$.

\subsection{Non-abelian bosonization of a SU(3) spin chain}
 The $SU(3)$ invariant
Hamiltonian $\tilde H_0$ can be solved exactly by the Bethe
ansatz\cite{uimin,sutherland}. The solution shows that the SU(3) spin
chain has 2 branches of excitations, with dispersion
$\epsilon_j(k)=\frac J 4 \frac{2\pi}{\sin (\pi j/3)}[\cos(\pi j/3-
|k|)-\cos (\pi j/3)],\; \; j=1,2$. These excitations are gapless,
and for $|k| \to 0$, one has $\epsilon_1(k)=\epsilon_2(k)\simeq
\frac{2\pi}{3} \frac J 4 \mid k \mid$, i.e. the dispersion relation
assumes at long wavelength a massless relativistic form. Accordingly,
the low energy, long wavelength excitations of the $SU(3)$ spin chain
can be bosonized. More precisely, these excitations are
described\cite{aff_wz,affleck_su(n)} by the 
SU(3) level 1 ($SU(3)_1$) Wess-Zumino-Novikov-Witten (WZNW)
model\cite{witten_wz}, perturbed by a marginally irrelevant $SU(3)$
invariant operator. A review of WZNW models can be found in
Ref. \onlinecite{tsvelikb}.
In Hamiltonian form, the $SU(3)_1$  model can be
written as: 
\begin{equation}\label{eq:wznw_model_general}
H_{WZNW}=\frac{2\pi}{3}\sum_{a=1}^{3^2-1}
:J^a_R(x)J_R^a(x):+:J^a_L(x)J_L^a(x): 
\end{equation}
where the right and left currents satisfy the following commutation
relations (Kac-Moody algebra at level 1):
\begin{equation}\label{eq:kac_moody_level1}
\lbrack J^{\alpha}_{R(L)}, J^{\beta}_{R(L)}\rbrack=if^{\alpha \beta \gamma}
\delta(x-y)J^{\gamma}_{R(L)}(y)+ \frac{i\delta_{\alpha \beta}}{2\pi}
\delta'(x-y).
\end{equation}
In Equation (\ref{eq:kac_moody_level1}), the $f^{\alpha \beta \gamma}$
are the structure constants of $SU(3)$. 
The central charge is $C=\frac{1 \times (3^2-1)}{3+1}=2$, indicating
that the $SU(3)_1$ WZNW model can be  described in terms of two free
bosonic fields.   
As mentioned above, the $SU(3)$ spin chain is described asymptotically
 by the $SU(3)_1$ model perturbed by a marginally irrelevant 
SU(3) invariant term,
\begin{equation}
\label{generalham}
H \to H_{WZW}+g_0 \int \frac{dx}{2\pi} \Phi_0(x),
\end{equation} 
where the marginal operator, $\Phi_0(x)=\sum_{\alpha=1}^{3} J_R^{\alpha}(x)J_L^{\alpha}( x)$,
couples the right and left currents.

 The finite size correction to the
ground state energy  of the $SU(3)$ chain  can be obtained
from the Bethe Ansatz solution. These corrections are logarithmic and
are in agreement with those obtained from the continuum Hamiltonian
(\ref{generalham}). 
This situation is very similar to the more familiar case of the
$SU(2)$ spin chain, which is described at low energy and long
wavelength by the marginally perturbed $SU(2)_1$ WZNW
model\cite{affleck_log_corr}.  
In general, the magnitude of $g_0$ cannot be obtained from the lattice
Hamiltonian in the case of a $SU(N)$ spin chain (see the discussion of
the case $N=2$ in Ref. \onlinecite{affleck_log_corr}).
This is even more problematic when one adds perturbations to the $SU(3)$
invariant spin chain. 
Another difficulty is that these perturbations 
 are not small in our case and strictly speaking 
cannot be treated in perturbation theory.
However, in one dimension weak  and strong
coupling behavior are often continuously
connected\cite{ziman_spin3/2,schulz_hubbard_exact,%
schulz_houches,kawakami_bethe_U<0} so that a weak coupling analysis
can provide very valuable information on the qualitative physics at
strong coupling.
Therefore, if we can find a weak coupling model that is described by
 marginally perturbed $SU(3)_1$ WZNW model and if we add to it small
perturbations of the form (\ref{lamb67})--(\ref{lamb88}), we will be
able to make a reasonable guess of the low energy long wavelength
continuum theory associated with the Hamiltonian $H_{\text{eff.}}$. 
By analogy with the Heisenberg model, we expect that the 
difference between the weak and the strong coupling regime will reduce to a
renormalization of some parameters of the effective low energy
theory. For non-integrable models, these parameters can be obtained
numerically by calculating  \emph{thermodynamic} quantities via
exact diagonalization methods
\cite{hayward_2chain,mila_zotos,ogata_tj}. 

In our case, it is not difficult to see that the spin sector of the
$SU(3)$ Hubbard model \cite{affleck_su(n),assaraf_su(n)} is a good candidate for a
weak coupling model. This model is defined by the Hamiltonian:
\begin{equation}
H=-t \sum_{i,n=1,2,3}(c^\dagger_{i+1,n}c_{i,n} +
\text{h.c.}) + U\sum_{i,n \ne m} n_{i,n} n_{i,m}
\end{equation}
where $c_{i,n}$ annihilates a fermion of flavor $n \in
[1,2,3]$ on site $i$, and $n_{i,n}=c^\dagger_{i,n}c_{i,n}$.
The basic idea is that, starting from the
lattice Hamiltonian of the $SU(3)$ Hubbard model it is possible to
take the continuum 
limit and then separate the  spin excitations from the
 charge excitations by means of weak coupling bosonization. 
In the strong coupling limit, $U \to \infty$, a constraint of  one
fermion per site is imposed,

\begin{equation}
\label{constraint}
\sum_{n}c^{\dagger}_{i,n}c_{i,n}=1.
\end{equation}
 With one fermion
per site, the charge degrees of freedom are frozen out and one is left 
only with SU(3) spin degrees of freedom.

Second order perturbation theory in $t$ then shows that the model can be mapped
onto an isotropic SU(3) spin chain with the lattice SU(3) spin operators
\begin{equation}\label{eq:abrikosov_su(3)}
\Lambda_i^{\alpha}=\sum_{n,m}
c^{\dagger}_{i,m}\lambda^{\alpha}_{n,m}c_{i,n},
\end{equation}
under the constraint (\ref{constraint}).
 Under the hypothesis of
continuity, the same $SU(3)_1$ field theory should describe the weak
and strong coupling limits in the spin sector. The difference between
the weak and strong coupling limit corresponds to
 the disappearance of the charge sector.  
This reduction of the number of degrees of freedom
can be obtained in a consistent way by treating the constraint
(\ref{constraint}) within the
effective theory \cite{itoikato,itoimukaida}.

Thus, our strong coupling theory is the spin sector of the $SU(3)$
Hubbard model with a filling of one fermion per site and $U\to \infty$. 
Let us discuss the weak coupling regime. The constraint
(\ref{constraint}) sets the Fermi momentum  at
$k_F=\frac{\pi}{3}$ for the three fermion flavors. Since we are
interested in low-energy, long wavelength properties, 
we linearize the spectrum for each flavor around the two Fermi points and 
introduce
the right and left moving fermion modes in the continuum limit, 
\begin{equation}
c_{i,n}^{\dagger} \simeq \sqrt{a}(e^{ik_Fx}\psi^{\dagger}
_{L,n}(x)+e^{-ik_Fx}\psi^{\dagger}_{R,n}(x)),
\end{equation}

\noindent where $x=ia$ and $a$ is the lattice spacing. 

For $U=0$, the linearized Hamiltonian is:
\begin{equation}
H_{\text{linearized}}=-\imath v_F \sum_n \int dx (\psi^\dagger_{R,n}\partial_x \psi_{R,n} - \psi^\dagger_{L,n}\partial_x \psi_{L,n}).
\end{equation}
This Hamiltonian is conformal invariant and 
can be rewritten in terms of the right and left
charge currents $J_{R(L)}=\sum_n \psi^\dagger_{R(L),n}\psi_{R(L),n}$,
and the eight SU(3) spin currents (right and left) $J^a_{R(L)}=\sum_n
\psi^\dagger_{R(L),n}\frac{\lambda^a_{n,m}}{2} \psi_{R(L),m}$. 
One thus separates the charge and spin sectors:
\begin{equation}
H=H_{\text{charge}}+H_{\text{spin}}
\end{equation}  
where the charge sector,
\begin{equation}
H_{\text{charge}}=v_F \int dx :J_R(x)J_R(x): + :J_L(x)J_L(x):
\end{equation}
and the spin sector is again described by the $SU(3)_1$ model discussed earlier,
\begin{equation}\label{ssuu}
H_{\text{spin}}=v_F \sum_a \int dx :J^a_R(x)J^a_R(x): + :J^a_L(x)J^a_L(x):
\end{equation}

The charge currents satisfy $U(1)$ Kac-Moody algebra, whereas the spin
currents satisfy the $SU(3)_1$ Kac-Moody algebra as can be checked
explicitly. When the interaction is weakly turned on,
$U/t\ll 1$, it
does not break spin charge separation but induces a $g_0 \propto
U$ term\cite{affleck_su(n)}. 

We  discussed thus far non-abelian bosonization in order
to stay close to the literature on $SU(3)$ spin chains. However, an
abelian bosonization approach to the isotropic $SU(3)$ spin chains
starting from the $SU(3)$ Hubbard model is perfectly
feasible. Such an
approach has been introduced for isotropic SU(N) spin chains 
in Ref. \onlinecite{assaraf_su(n)}. It is
outlined in  appendix \ref{app:abelian_hubbard_su(3)}. 
In fact, for the rest of this section, we shall employ abelian
bosonization  because it renders 
 the calculation of
correlation functions  extremely easy even when the $SU(3)$ symmetry is
explicitly broken.  

\subsection{Abelian bosonization approach}

Abelian
bosonization gives the following Hamiltonian for an $SU(3)$ invariant 
spin chain
(or the spin sector of the $SU(3)$ Hubbard model):
\begin{eqnarray}
H_{su(3)} & = &\int \frac{dx}{2\pi} u\left[ (\pi \Pi_a)^2 +(\pi \Pi_b)^2 +
(\partial_x \phi_a)^2 +(\partial_x \phi_b)^2 \right]\nonumber \\ 
& + & \frac{2Ua}{(2
\pi a)^2}\int dx (\cos \sqrt{8}\phi_a +\cos \sqrt{2}(\phi_a +
\sqrt{3}\phi_b) + \cos \sqrt{2}(\phi_a -\sqrt{3}\phi_b) ) \nonumber \\
& - &  Ua \int
\frac {dx}{\pi^2} \left[ (\partial_x \phi_a)^2 + (\partial_x \phi_b)^2
\right].
\end{eqnarray}
A derivation can be found in Appendix
\ref{app:abelian_hubbard_su(3)}.
The
free term corresponds to Eq.(\ref{ssuu}).

Under renormalization,  $H_{su(3)}$ flows 
 to a fixed point Hamiltonian\cite{assaraf_su(n)}:
\begin{equation}
H^* = \int \frac{dx}{2\pi} u^*\left[ (\pi \Pi_a)^2 +(\pi \Pi_b)^2 +
(\partial_x \phi_a)^2 +(\partial_x \phi_b)^2 \right],
\end{equation}
where $u^*$ is given by the Bethe Ansatz as $u^*=\frac{2\pi}{3} \frac
J 4 $. 
One can check using the expressions (\ref{eq:lambda_operators})
 that this leads to a scaling dimension of $1$ for the
uniform component of $\Lambda^{\alpha}(x) \simeq a^{-1}\Lambda_i^{\alpha},\; x=ia $ ($\alpha=1 \cdots 8$), 
and $2/3$ for the $2\pi/3$ component (see Eq. (\ref{eq:lambda_operators})).
 These scaling dimensions coincide with
those obtained from non-abelian bosonization\cite{aff_wz,assaraf_su(n)}.

Turning now to the $SU(3)$ symmetry breaking terms, we find that 
in the abelian bosonization representation they take the following form:
\begin{eqnarray}
\tilde H_1 & = &  \int \frac {dx}{\pi^2} \left[ -\frac {qa}{4} (\pi
 \Pi_a)^2 - \frac {qa} 2(\partial_x \phi_a)^2 +
 \frac {qa}{12} (\partial_x \phi_b)^2 \right] +
 \frac{2qa}{(2 \pi a)^2}\int dx \cos \sqrt{8} \phi_a + \frac{\sqrt{3}
 qa}{(2\pi)^2} \int dx \partial_x \phi_b, \label{h11}\nonumber \\
\tilde H_2& = & \int \frac{dx}{\pi^2} \left[ \frac{5ua}{2}
(\partial_x \phi_a)^2 + \frac{ua}{6} (\partial_x \phi_b)^2 \right] +
\frac{2ua}{4(\pi a)^2} \int dx \cos (\sqrt{8}\phi_a) + \int
 \frac{\sqrt{3}u}{(2\pi)^2} \partial_x \phi_b ,\label{h2}\\
\tilde H_3 & = &  \int \frac{dx}{\pi^2} \left[
\frac{u'a}{6} (\partial_x \phi_a)^2 + \frac{5u'a}{2}(\partial_x \phi_b)^2 \right]
 -\frac{2u'a}{3(2\pi a)^2} \int dx \cos (\sqrt{8}\phi_a)-
\nonumber \\
 &  &\; \; \; \; \; -\frac{4u^\prime a}{3(\pi a)^2} \int dx \cos \sqrt{2} \phi_a
 \cos \sqrt{6} \phi_b + \frac{u^\prime \sqrt{2}}{3 \pi^2} \int
 \partial_x \phi_b,\label{h3}\\
\tilde H_h&=& -(h-h_c)
\int dx \frac{\sqrt{2}}{\pi} (\partial_x \phi_b).\label{hh}
\eea
The physical interpretation of the terms proportional to $\partial_x
 \phi_b$ is very simple. The bosonized Hamiltonian is derived under
 the assumption that the magnetization per triangle is close to
 $5/(6a)$. When the magnetization per triangle is \emph{exactly}
 $5/(6a)$ the terms  $\partial_x \phi_b$ do not appear in the
 Hamiltonian. Therefore, the presence in the Hamiltonian of such terms
  means that the magnetic field
 needed to impose a magnetization of $5/(6a)$  per triangle  is
 renormalized away from its bare value. Also, since 
 the Hamiltonian preserves the
 symmetry between $+$ and $-$ chiralities, it is invariant under the
 transformation $\pi_a \to -\pi_a$, $\phi_a \to -\phi_a$. In
 particular, this precludes the  terms $\partial_x \phi_a$
 in the Hamiltonian.  

Assembling all terms we finally have the following
 field theory describing the spin sector of the SU(3) Hubbard model in
 the presence of  symmetry breaking perturbations,

\begin{eqnarray}
H & = & v_F \sum_{i=a,b} \int \frac{dx}{2\pi} \lbrack
((\pi \Pi_i)^2+(\partial_x \phi_i)^2\rbrack+
+\frac{2g_1}{(2\pi a)^2}\int dx \cos \sqrt{8} \phi_a(x)\nonumber \\
& + &
\frac{4g_2}{(2\pi a)^2}\int dx \cos \sqrt{2} \phi_a(x)\cos \sqrt{6}
\phi_b(x) +  \frac{g_4}{\pi^2} \int dx (\partial_x \phi_a)^2 +
\frac{g_5}{\pi^2} \int dx (\partial_x \phi_b)^2 +\frac{h}{\pi}\int
dx \partial_x \phi_b
\label{htott}
\end{eqnarray}

\noindent with $v_F=2ta\sin (k_Fa)=\sqrt{3}ta$, 
$h=(\frac{\sqrt{3}}{4\pi}u+\frac{\sqrt{2}}{3\pi}u'+
\frac{\sqrt{2}}{2\pi}q)$, and $t$ the hopping
amplitude. In our units, $t=1$.

The notations can be made more compact by introducing the vectors
$\vec{\phi}=(\phi_a,\phi_b)$ and $\vec{\alpha}_1=(1,0)$ , 
$\vec{\alpha}_2=(1/2,\sqrt{3}/2)$  and
$\vec{\alpha}_3=(1/2,-\sqrt{3}/2)$, where,

\begin{eqnarray}
K_a &=&\left(\frac{1-\frac{qa}{2\pi v_F}}{1-\frac{Ua}{\pi
v_F}-\frac{qa}{\pi v_F}+\frac{u'a}{3\pi v_F}+\frac{5ua}{\pi
v_F}}\right)^{1/2}\nonumber \\
u_a &=&v_F \left[\left(1-\frac{qa}{2\pi v_F}\right)\left(1-\frac{Ua}{\pi
v_F}-\frac{qa}{\pi v_F}+\frac{u'a}{3\pi v_F}+\frac{5ua}{\pi
v_F}\right)\right]^{1/2}\nonumber \\
K_b &=&\left( \frac 1 {1-\frac{Ua}{\pi v_F}+\frac{5u'a}{\pi v_F}+\frac{ua}{3 \pi
v_F} +\frac{qa}{6\pi v_F}}\right)^{1/2}\nonumber \\
u_b &=&v_F\left(1-\frac{Ua}{\pi v_F}+\frac{5u'a}{\pi v_F}+\frac{ua}{3 \pi
v_F} +\frac{qa}{6\pi v_F}\right)^{1/2}.
\end{eqnarray}

The Hamiltonian can then be rewritten\cite{note_fusion_2d} as,
\bea
H=\int \frac{dx}{2\pi}\left[u_a K_a (\pi
\Pi_a)^2+\frac{u_a}{K_a}(\partial_x \phi_a)^2 + u_b K_b(\pi
\Pi_b)^2 +\frac{u_b}{K_b}(\partial_x \phi_b)^2 \right]+
\sum_{i=1}^3 \frac{2 g_i}{(2\pi a)^2} \int dx \cos (\sqrt{8} \vec \alpha_i. \vec \phi) \nonumber
\eea
The interactions $(\partial_x \phi_{a,b})^2$ can render the
marginal operators $\cos (\sqrt{8}\phi_a)$,
$ \cos (\sqrt{2}\phi_a \pm  \sqrt{6}\phi_b)$ marginally relevant and
cause the opening of a gap.
In such a case, the  low energy properties of the system cannot be
described by two massless bosons. One can have either a massive and a
massless boson or two massive bosons.
This depends on the coupling constants $u$, $u'$, $m$, $q$, and the
magnetic field $h$. In order to explore this possibility in more
 detail, one has to use renormalization group equations. This is the
 subject of the forthcoming sections.

\section{The renormalization group flow in zero magnetic field}\label{RG}

In this section, we discuss the flow of the renormalization group
equation and the phase diagram that results from it. Qualitatively,
the renormalization group equations are similar to the
Kosterlitz-Thouless renormalization group
equations\cite{kosterlitz_renormalisation_xy,jose}. We expect
therefore to obtain a gapless phase  corresponding to the
 flow  to a fixed hypersurface of the
6-dimensional space of coupling constants and one (or possibly many)
gapped phase where the coupling constants flow to infinity.
We also expect that the phase transition is of infinite
order\cite{kosterlitz_renormalisation_xy}.
Our task is therefore to determine the initial conditions and follow
 the  flow. This will allow us to conclude on the nature of
the ground state of the anisotropic SU(3) chain.

 A straightforward application to the Hamiltonian (\ref{htott}) 
of the standard method
\cite{jose,giamarchi_logs} would be
inconvenient since one needs to  expand to
third order of correlation functions in order to get the full one loop
RG equations\cite{nelson_fusion_2d}.  
We will use instead   operator product
expansion (OPE)
techniques\cite{kadanoff_gaussian_model,knops_sine-gordon}. 
In our case, the  algebra of operators $(\partial_x
\phi_a)^2$, $(\partial_x \phi_b)^2$ and  $\cos (\sqrt{8} \vec \alpha_i .\vec
\phi)$ close under OPE (for details see Appendix \ref{app:derivation_ope}).
In particular,
\begin{eqnarray}
\label{opem}
\cos (\sqrt{8}\vec \alpha \vec \phi(x,\tau))\cos (\sqrt{8}\vec \alpha
\vec  \phi(0,0)) & \simeq & \frac{-2a^4}{(x^2+(u\tau)^2)^2} \left
[ \sum_{p=a,b}   (\alpha_i^p)^2
(x^2 (\partial_x \phi_p)^2 + \tau^2 (\partial_\tau \phi_p)^2 + 2 x\tau
\partial_x \phi_p \partial_\tau \phi_p \right],\nonumber \\ 
\cos(\sqrt{8}\vec \alpha \vec \phi(x,\tau))\cos(\sqrt{8}\vec \beta
\vec \phi(0,0))& & \simeq \frac 1 2
\left[ \left(\frac {a^2} {x^2+(u\tau)^2}\right)^{-K \vec \alpha.\vec
\beta}\cos \sqrt{8}(\vec \alpha +\vec \beta).\vec
\phi(0,0) + \right.\nonumber \\
&  & \;\;\;\;\left. \left(\frac{a^2}{x^2+(u\tau)^2}\right)^{K \vec \alpha . \vec \beta}
\cos \sqrt{8}(\vec \alpha -\vec \beta).\vec \phi(0,0) \right],
\nonumber 
\end{eqnarray}
lead to  the following
 RG equations (see appendix \ref{app:derivation_ope}):

\begin{eqnarray}
\frac{dy_1}{dl} & = & 2y_1y_4 - y_2^2/2 \nonumber \\
\frac{dy_2}{dl} & = & (1/2 y_4+3/2 y_5)y_2 - y_1y_2/2 \nonumber \\
\frac{dy_4}{dl} & = & y_1^2/2+y_2^2/4 \nonumber \\
\label{rge}
\frac{dy_5}{dl} & = & 3/4 y_2^2,
\end{eqnarray}

\noindent where we have denoted $y_i=g_i/\pi v_F$, with $v_F$ the Fermi
velocity. Note that is a fixed surface
$(g_1,g_2)=(0,0)$, because of three truly marginal operators
$\partial_x \phi_a \partial_x \phi_b$, $(\partial_x \phi_a)^2$ and $
(\partial_x \phi_b)^2$.

An alternative approach based on non-abelian bosonization can be used. In 
this approach, having expressed the
Hamiltonian in terms of products of right and left moving currents
$J_R^a J_L^a$, an operator product expansion for 
currents is derived\cite{itoi_rg_calculation}. Such an approach leads
to the same RG equations as the Abelian bosonization
approach.

 The initial values of the running coupling constants (at the cut off 
 scale $a$)
for the spin sector of the SU(3) Hubbard model perturbed by $\tilde
H_{1,2,3}$ are given by,

\begin{eqnarray}
y_1(a)& = & \frac{g_1(a)}{\pi v_F}=(Ua +qa - \frac{u'a}{3} +ua)/\pi
v_F)
\nonumber \\
y_2(a)& = & \frac{g_2(a)}{\pi v_F} = (Ua +\frac{u'a}{3})/\pi v_F
\nonumber \\
y_4(a) & = & \frac{g_4(a)}{\pi v_F} = (-\frac{Ua}{2} +\frac{5ua}{2}+
\frac{qa}{2} + \frac{u'a}{6})/\pi v_F \nonumber \\
\label{starp}
y_5(a) & = &  \frac{g_5(a)}{\pi v_F} = (-\frac{Ua}{2} +\frac{ua}{6} +
\frac{5u'a}{2} + \frac{qa}{12})/\pi v_F.
\end{eqnarray}
In the expression for the initial coupling constants (\ref{starp}) we
have $u=-J/4$, $q=-J/12$, $u'=-5J/36$, and we assume $J,U \ll t$. 
Hereafter, we choose $J=4$
and put $v_F$ equal to unity, thus the numerical starting values are
given by

\begin{eqnarray}\label{eq:ini_cond}
y_1 & = & -0.365467+y_0 \nonumber \\
y_2 & = & -0.0589463 + y_0 \nonumber \\
y_4 & = & -0.878299 -y_0/2 \nonumber \\
y_5 & = & -0.5039907 -y_0/2.
\end{eqnarray}
Here $y_0 \propto U$.
These values are not small, so that the one loop RG equations are not
valid. However, numerically solving these RG equations with initial
conditions (\ref{eq:ini_cond}) shows that they
flow to a fixed point on the surface $(g_1,g_2)=(0,0)$ 
for any $y_0 \in [0,1]$ (see Figs. \ref{fig:rgflow_u=0} and
\ref{fig:rgflow_u=1}). At this fixed point, one has a renormalized
Hamiltonian with:
\begin{equation}
H=\int \frac {dx} {2\pi} \left[ u_a^* K_a^* (\pi \Pi_a)^2 +
\frac{u_a^*}{K_a^*} (\partial_x \phi_a)^2 + u_b^* K_b^* (\pi \Pi_a)^2 +
\frac{u_b^*}{K_b^*} (\partial_x \phi_a)^2 \right] 
\end{equation} 
This proves that certainly at weak coupling the long distance
properties of the system are described by a two component Luttinger
liquid. At strong coupling, i. e. in the case of the spin tube, this
is only an indication that the two component Luttinger liquid is
possible. In order to give a definitive proof, one should prove that
there is no singularity in the ground state energy as couplings
increase. Comparing Fig.~\ref{fig:rgflow_u=0}
and~\ref{fig:rgflow_u=1}, one can see that the magnitude of the fixed
point values of $g_4$ and $g_5$ depends on the strength of the
marginal SU(3) symmetric interaction $y_0$. 
This fact, combined with the fact that the RG equations are only valid
at weak coupling precludes the use of the RG to give an accurate
estimate of $K_a^*$ and $K_b^*$. However, one can still determine from
the RG equations whether these quantities are larger or smaller than
one.    
Although we stress that these figures should not be taken too
seriously, 
we find using the expressions of $K_a$ and $K_b$ as a function of
$g_4$ and $g_5$ $K_a^*=1.9$ and $K_b^*=1.5$ i.e. both are larger than
1. Concerning the question 
whether the two component Luttinger liquid persists at
large coupling, we can remark that the deviation from isotropy in our
case makes the interaction between the SU(3) spins less
antiferromagnetic. It is well known that in the case of the XXZ chain,
reducing the antiferromagnetic character of the spin spin interaction
(i.e. working at $J_z <J$ ) prevents the formation of a gap
\cite{haldane_xxzchain}. Therefore, it seems likely that no gap would
develop in the spectrum.   
 To test our conjecture, numerical work, especially
calculation of $K_a^*,K_b^*$ by exact diagonalization would prove
very valuable. 

The existence of a two component Luttinger liquid phase has important
consequences. In particular, it implies a non-zero magnetic
susceptibility $\chi \propto K_b/u_b$, and a $T$ linear specific heat
of the form:
\begin{equation}
\label{eq:specific_heat}
C=\frac {\pi T}{6 u_a} +\frac{\pi T}{6 u_b} .
\end{equation}

The calculation of the correlation functions and NMR relaxation rate
are postponed to section \ref{correlations}.

\section{Strong magnetic field case: fixed point Hamiltonian and
correlation functions}\label{correlations}

\subsection{ Generic magnetic field}

\subsubsection{Renormalization group flow under a magnetic field}
Till now, we haven't taken into account the terms associated with the
magnetic field $h_b$, which can be treated as a
perturbation having fixed the external magnetic field at
$h_c=3/2J_{\perp}$. To see if the flow remains unchanged in
this case, let us reobtain the renormalization group equations with
finite $h$. The simplest way to address this problem is to perform a
Legendre transformation\cite{giamarchi_spin_flop} on the Hamiltonian
(\ref{htott}). The non-zero
average value of the field $\phi_b$ due to the finite magnetization can
be eliminated by a simple shift of the  $\phi_b$ fields,
i.e. $\phi_{b}=\phi_{b}- \pi m_{b}x$, where $m_b=-\langle \partial_x
\phi_b \rangle /\pi $. One has the relation between $m_b$ and the
magnetization :
\begin{equation}
\label{eq:magnetization_mb}
m_b=-\sqrt{\frac 3 2}\left( \frac{\langle S^z \rangle}{a} -\frac 5 {6a}\right)
\end{equation}

The cosine terms, however,
are not invariant under this shift and the renormalization group
equations (\ref{rge}) for the couplings $g_1$, $g_2$ and $g_3$,
for a change of the length scale $a\rightarrow
e^{l} a$, now become (the details of the calculation can be found in
the appendix \ref{rghh}):

\begin{eqnarray}\label{betafh}
\frac{dy_1}{dl} & = & 2y_1y_4 -y_2^2 J_0(\pi m_b(l) a \sqrt{3})/2 \nonumber \\
\frac{dy_2}{dl} & = & (1/2 y_4+3/2 y_5)y_2 -y_1y_2 J_0(\pi m_b(l) a \frac{\sqrt{3}}2)/2 \nonumber \\
\frac{dy_4}{dl} & = & y_1^2/2+y_2^2 J_0(\pi m_b(l) a \frac{\sqrt{3}}2)/4 \nonumber \\
\frac{dy_5}{dl} & = & 3/4 y_2^2 J_0(\pi m_b(l) a \frac{\sqrt{3}}2),
\end{eqnarray}
\noindent where $J_0$ is the Bessel function that results from the use
of a sharp cutoff in the real space. One also has
$m_b(l)=m_b(0)e^l$. One can check that setting $m_b(0)=0$ one recovers
the equations (\ref{rge}).
If the RG equation for the magnetization is trivial, 
the magnetic field on the other hand, satisfies a non-trivial RG
equation:
\begin{equation}
\frac{dh_b}{dl}=h_b+\frac{\sqrt{3}}{a \sqrt{8}}y_2^2 J_1(\pi \sqrt{6}
m_b(l) a).
\end{equation}

Let us discuss qualitatively the physics predicted by the
Eqs. (\ref{betafh}). One sees rather easily that for $m_b (l) a \ll
1$, the Bessel functions tend to zero, so that one is left with a
Sine-Gordon renormalization group equation for $y_1,y_4$. 
Compared to the case of zero magnetization, we see that $y_4$ is more
negative and $y_1$ is smaller in absolute value. Therefore, we expect
that $y_1$ will be even more irrelevant in the presence of the finite
magnetization. 
We conclude that the presence of a non-zero magnetization does not
affect the two component Luttinger liquid behavior.   
The crossover scale can be roughly estimated as:
\begin{equation}
\label{lstar}
l^{\star} \simeq ln(\frac{v_F}{m_b a}).
\end{equation}
At this crossover scale, the flow of $y_5$ is completely
cut. This implies a variation of $K_a,K_b$ with the magnetization. 

 At the value
of $l$ given by (\ref{lstar}) the magnetic energy is of
the order of energy cut-off, therefore  the magnetic field term
cannot be treated as a perturbation. When the initial magnetization
goes to 
infinity the renormalization is stopped for smaller and smaller $l$. 
The coupling constants $g_1,g_2,g_3$ then become
zero, while $g_4,g_5$ assume the values they have at the scale
$l^*$. Returning to the Hamiltonian (\ref{htott}), we see that it
becomes a quadratic Hamiltonian.

\subsubsection{Fixed point Hamiltonian}
Following the the preceding discussion, we conclude that
the asymptotic behavior of the three
chain system under a magnetic field is governed by the  Hamiltonian:

\bea
H^{\star}= \int \frac{dx}{2\pi} v_F \left[ (\pi \Pi_a)^2 + (\partial_x
\phi_a)^2 + (\pi \Pi_b)^2 +(\partial_x \tilde \phi_b)^2 \right]
+\frac{g^*_4}{\pi^2}\int dx (\partial_x \phi_a)^2
+\frac{g^*_5}{\pi^2}\int dx (\partial_x \phi_b)^2  \nonumber
\eea
\noindent where
$g_{4,5}^*$ are functions of the magnetic field.
The field $\tilde \phi_b$ is related to $\phi_b$ in the following way:
\begin{equation}
\phi_b =  \tilde \phi_b+ \pi (m-\frac 5 {6a})\sqrt{\frac 3 2} x,
\label{shift}
\end{equation}
\noindent while the dual fields $\theta_a$ and $\theta_b$ are
not shifted. This condition guarantees that $\tilde \phi_b$ satisfies
periodic boundary conditions.

 The fixed point Hamiltonian  can be rewritten:
\begin{equation}
\label{hcrittt}
H^{\star}= \int \frac{dx}{2\pi} \left[ u^*_a K^*_a(\pi
\Pi_a)^2+\frac{u^*_a}{K^*_a}(\partial_x \phi_a)^2\right] +
\int \frac{dx}{2\pi} \left[ u^*_b K^*_b(\pi
\Pi_b)^2+\frac{u^*_b}{K^*_b}(\partial_x \tilde \phi_b)^2\right],
\end{equation}

\noindent where $u^*_iK^*_i=v_F; \; i=a,b$ and
$u^*_{a,b}/K^*_{a,b}=v_F +2 g_{4,5}^*/\pi$. 
Both the velocities of excitations, $u_i$,
and the compactification radii, $K_i$, depend on the magnetic field $h$
 through 
$g_i^{\star}(h)$.
Therefore, the low energy properties of the system are described by
two decoupled $c=1$ conformal field theories with velocities and
compactification radii depending on the applied magnetic field.

This is valid at the level of perturbation theory for the
spin sector of the SU(3) Hubbard model. However, we are actually
interested in the SU(3) anisotropic spin chain for which perturbation
theory does not apply. In the latter case, we
expect, relying on the continuity between the weak and the strong
coupling regime, that the anisotropic SU(3) spin chain under magnetic
field will also be described by a two component Luttinger liquid. 
However, the velocities and compactification radii cannot be
 obtained by perturbation theory techniques.
Nevertheless, it is known that the velocities and compactification
radii can be obtained by calculating only thermodynamic quantities
using, for instance, exact diagonalization
techniques\cite{ogata_tj,mila_zotos}.
The problem of the determination of these exponents
in terms of measurable thermodynamic
quantities in the specific case of the anisotropic SU(3) spin chain
 will be discussed in  Appendix
\ref{app:exposants_bosonized}. The knowledge of the exponents then
permits the calculation of correlation functions. This is the subject
of the forthcoming section.  

\subsubsection{Correlation functions}

In this section, we want to calculate the three Matsubara 
correlation functions:
\bea
\chi_{zz}(x,\tau)&=&\langle T_\tau S^z(x,\tau) S^z(0,0) \rangle \label{eq:totspin_corr}\\
\chi_{+-,p}(x,\tau)&=& \langle T_\tau S_p^+(x,\tau) S_p^-(0,0) \rangle \label{eq:locspin_corr}\\
\chi_{zz,p}(x,\tau)&= &\langle T_\tau S_p^z(x,\tau) S_p^z(0,0) \rangle \label{eq:locspin_corr_z},
\eea
where $p=1,2,3$ is a chain index. 
The first correlation function is useful for neutron scattering
experiments, whereas the  correlation functions
(\ref{eq:locspin_corr})
 are useful for
the calculation of NMR relaxation rates.
The Matsubara correlation functions in Fourier space are given by: 
\begin{equation}
\label{susc}
\chi_{ij}(q,\omega_n,T)= \int_0^\beta \;d\tau dx \exp^{\imath(\omega_n \tau - q x)} 
\langle T_\tau \lbrack S^i(x,\tau),S^j(0,0)\rbrack \rangle_T,
\end{equation}
from which the finite temperature correlations are obtained by the
analytic continuation $\imath \omega_n \to \omega +\imath 0_+$.
We will first concentrate on  the $T=0$
calculation, then explain how to extend the calculation to finite
temperature.

We begin with the calculation of $\chi_{zz}$. Using the equation
(\ref{siz}) we have: 
\begin{equation}
\chi_{zz}=\frac 1 3 \langle T_\tau \Lambda^8(x,\tau) \Lambda^8(0,0)
\rangle +(\langle S^z\rangle)^2 .
\end{equation}

Using the bosonized expressions of the SU(3) spins,
Eq. (\ref{eq:lambda_operators}), and the usual expression for the bosonized 
correlation
functions\cite{schulz_moriond}, we obtain\cite{note_logs_1}:
\begin{eqnarray}
\chi_{zz}& = & (\langle S^z\rangle)^2 +\frac {K_b} {3\pi^2} 
\frac{(u_b\tau)^2-x^2}{(x^2+(u_b \tau)^2)^2} +
  \frac {e^{\imath (\frac {2 \pi
x}{3a} -\pi (m -\frac 5 {6a}))x}} {6 (\pi a)^2}  \left(\frac
{a^2}{x^2 +( u_a \tau)^2}\right)^{\frac {K_a}{6}}  \left(\frac
{a^2}{x^2 +( u_b \tau)^2}\right)^{\frac {K_b}{2}}  \nonumber \\  
&  &\;\;\;+ \frac {e^{\imath (\frac {2 \pi
x}{3a} +2\pi (m -\frac 5 {6a}) )x}} { 3(\pi a)^2}  
 \left(\frac
{a^2}{x^2 +( u_b \tau)^2}\right)^{\frac {2K_b}{3}}  +\text{c.c},
\end{eqnarray}
where $m=\langle S^z \rangle /a$. 

Turning to $\chi_{+-,p}$, it is easily seen using Eq. (\ref{si}) that 
it is independent of
$p$ and equal to:
\bea
\chi_{+-,p}(x,\tau)=\frac 1 {12} \left[ \langle T_\tau (\Lambda^1 +\imath
\Lambda^2)(x,\tau)(\Lambda^1 -\imath \Lambda^2)(0,0)\rangle + \langle
T_\tau (\Lambda^4 + \imath \Lambda^5) (x,\tau) (\Lambda^4 - \imath
\Lambda^5)(0,0) \rangle \right] \nonumber
\eea
Similarly, $\chi_{zz,p}$ is independent of $p$ (see Eq. (\ref{si}) 
and the contribution
not already included in $\chi_{zz}$ is of the form:
\begin{equation}
\langle T_\tau (\Lambda^6 + \imath \Lambda^7) (x,\tau) (\Lambda^6 - \imath
\Lambda^7)(0,0) \rangle.
\end{equation}
The expressions of the required correlators are obtained as:

\begin{eqnarray}\label{eq:general_correlation}
 \langle T_\tau (\Lambda^n + \imath \Lambda^{n+1}) (x,\tau)(\Lambda^n - \imath
\Lambda^{n+1})(0,0) \rangle = \frac 2 {(\pi a)^2}\left[\left(\frac
{a^2}{x^2 +(u_a \tau)^2}\right)^{\nu_{n,1}} \left(\frac{a^2}{x^2 +(u_b
\tau)^2}\right)^{\nu_{n,2}} \right. \nonumber\\ 
\left. \times  \cos (Q_n x + \Phi_n(\tau/x)) +  \left(\frac
{a^2}{x^2 +(u_a \tau)^2}\right)^{\eta_{n,1}} \left(\frac{a^2}{x^2 +(u_b
\tau)^2}\right)^{\eta_{n,2}}
  \cos (\frac {2\pi}{3a} + Q_n^\prime x +
\Psi_n(\tau/x)) \right] 
\end{eqnarray}
where $n=1,4,6$.
The exponents are given by:

\begin{eqnarray}
\nu_{1,1}&=&\frac 1 {2K_a}+ \frac {K_a} 2 \; \nu_{1,2}=0 \\
\eta_{1,1}&=&\frac 1 {2K_a} \; \eta_{1,2}=\frac {K_b} 6 \\
\nu_{4,1}&=&\nu_{6,1}=\frac 1 {8 K_a} + \frac {K_a} 8 \;
\nu_{4,2}=\nu_{6,2}= \frac 3 {8K_b} + \frac {3 K_b} 8 \\
\eta_{4,1}&=&\eta_{6,1}=\frac 1 {8 K_a} + \frac {K_a} 8 \;
\eta_{4,2}=\eta_{6,2}=\frac 3 {8 K_b} + \frac {K_b} {24}.  
\end{eqnarray}
 It can be checked  that for
 $u_a=u_b$, $K_a=K_b=1$, one recovers the exponents of the isotropic SU(3) spin
 chain\cite{aff_wz}, namely  $\nu_{n,1}+\nu_{n,2}=1$ and
 $\eta_{n,1}+\eta_{n,2}=2/3$.
One also has:
\begin{eqnarray}\label{eq:q_vectors}
Q_1&=&0, \;\; \;Q_1^\prime = -\pi \left(m -
\frac{5}{6a}\right) 
\nonumber \\
Q_4&=&\frac 3 2  Q_1^\prime=-Q_6,\;\; \;\;\; Q_4^\prime = -\frac
{Q_1^\prime}{2}=-Q_6^\prime.
\end{eqnarray}
Recall $\langle S^z \rangle=5/6 +\sqrt{2/3} m_b$.
This allows the determination of all incommensurate modes. 
Finally, the functions:
\begin{eqnarray}
\Phi_1\left(\frac \tau x\right)&=&2\arctan \left(\frac{u_a \tau}
x\right), \; \; \Psi_1\left(\frac \tau x\right)=0 \nonumber \\
\Phi_4\left(\frac \tau x\right)&=&\Phi_6\left(\frac \tau x\right)=\frac
1 2 \arctan  \left(\frac{u_a \tau} x\right) +\frac 3 2 \arctan
\left(\frac{u_b \tau} x\right)  \nonumber \\
\Psi_4\left(\frac \tau x\right)&=&-\Psi_6\left(\frac \tau x\right)=\frac
1 2 \left[\arctan  \left(\frac{u_a \tau} x\right)-\arctan
\left(\frac{u_b \tau} x\right)\right]
\end{eqnarray}

All the preceding results are valid only at $T=0$. However, it is
useful also to
calculate the correlation functions  for $T>0$,
in particular in order to obtain NMR relaxation rates. 
To obtain the finite temperature Matsubara correlation functions,
 we can use a conformal transformation since we have two decoupled 
$c=1$ conformal field theories. The explicit expression of this
transformation is:
\begin{equation}\label{eq:conformal_mapping}
x+\imath u_i \tau \to \beta u_i \sinh \left( \frac  {2\pi (x+ \imath
u_i \tau)} {\beta u_i} \right),
\end{equation}  
where $i=a,b$. 
Therefore, to obtain\cite{note_logs_2} the finite temperature Matsubara correlation
functions, one has to make the substitution:
\begin{eqnarray}
\label{eq:conf_substitution}
x^2+(u_i \tau)^2 &\to& (\beta u_i)^2 \left[ \cosh^2 \left(\frac {2\pi x} {\beta
u_i}\right) - \cos^2 \left( \frac {2\pi \tau} \beta \right) \right]
\nonumber \\
\arctan \left(\frac{u_i \tau} x \right) &\to&
\arctan\left(\frac{\tan\left(2\pi \frac \tau \beta\right)}{\tanh \left(
2 \pi \frac x {\beta u_i}\right)}\right).
\end{eqnarray}

With the help of the above results for the spin-spin correlation
functions we can  evaluate the $T$ dependence of the NMR 
longitudinal relaxation rate  $T_1$, 
\begin{equation}
\frac{1}{T_1} \propto \lim_{\omega_n \rightarrow 0}\int_0^\beta d\tau
e^{\imath \omega_n \tau} \langle T_\tau S^+_p(0,\tau) S_p^-(0,0) \rangle_T.
\end{equation}
We find,
\bea
\frac 1 {T_1} \propto (a~ T^{\frac 1 {2K_a} +\frac{K_a} 2 -1} +b~ T^{\frac 1
{2K_a} +\frac {K_b} 6} +c ~T^{\frac{3 K_a+  K_b} {24} + \frac 1 {8 K_a} + \frac
3 {8 K_b} -1} +d ~T^{\frac 1 {8K_a} + \frac 3 {8K_b} + \frac {3 K_b +
K_a} 8 -1 } ) \nonumber
\eea
The  low temperature exponent is the smallest
 of the the four exponents above.

\subsubsection{Comparison with a spin-1 chain with biquadratic coupling}

In the case of a bilinear biquadratic spin-1 chain defined in
Eq. (\ref{eq:bilinear_biquad}), close to the
Uimin-Lai-Sutherland point ($\beta \simeq 1$), 
a mapping onto an anisotropic SU(3) spin
chain is also possible\cite{fath_biquadratic}. 
However, there are important differences.
First, the expression of the spin operators in terms of the
Gell-mann matrices is  different from the ones obtained in
the spin-tube case.
For the spin-1 case, one has:
\begin{eqnarray}\label{eq:spin1_lambda}
S_n^x& = & \frac{\sqrt{2}}{2}\left(\lambda_n^4+\lambda_n^6\right) \nonumber
\\
S_n^y& = & \frac{\sqrt{2}}{2}\left(\lambda_n^5-\lambda_n^7\right) \nonumber
\\
S_n^z& = &\lambda_n^3.
\end{eqnarray}
These expressions should be contrasted with Eqs.(\ref{si})--(\ref{siz}).
\noindent Although the expressions (\ref{eq:spin1_lambda})
 lead to incommensurate modes, the expression of the
correlation functions is different from the case of the spin tube.
Second, the expression of the Hamiltonian in terms of $\lambda$
matrices in the spin-1 case is different from the expression
(\ref{heff}). Namely, the Hamiltonian (\ref{eq:bilinear_biquad}) can
be rewritten in terms of Gell-Mann matrices as:

\begin{eqnarray}
\label{eq:bilin_biquad_su(3)}
H& = &\sum_i \left[\frac \beta 2 (\lambda^8_i \lambda^8_{i+1} +\lambda^1_i
\lambda^1_{i+1} + \lambda^2_i
\lambda^2_{i+1}) + (1-\frac \beta 2) \lambda_i^3
\lambda_{i+1}^3 \right.\nonumber \\
& +& \left. \frac 1 2 (\lambda^4_i \lambda^4_{i+1}+ \lambda^5_i \lambda^5_{i+1} +
\lambda^6_i \lambda^6_{i+1} + \lambda^7_i \lambda^7_{i+1}) + \frac
{1-\beta} 2 (\lambda^4_{i} \lambda^6_{i+1} +  \lambda^6_{i}
\lambda^4_{i+1}- \lambda^5_{i} \lambda^7_{i+1} -\lambda^7_{i}
\lambda^5_{i+1})\right] 
\end{eqnarray}

Finally, for $\beta<1$ the spin-1 bilinear biquadratic chain has a gap
and the two component Luttinger liquid can only be observed for a
large enough applied magnetic field.

Nevertheless, the two problems have in common the
presence of a gapless two component Luttinger liquid ground
state\cite{fath_biquadratic}, and
the formation of incommensurate modes under a magnetic field, so that
loosely speaking they belong to the same universality class.
This can be understood as a consequence of the fact that both models
can be related to anisotropic SU(3) spin chains. 
One should note that the formation of incommensurate modes in the
presence of the magnetic field in the spin-tube
is not related to the presence of \emph{gapped} incommensurate modes
in the bilinear-biquadratic spin-1 chain
\cite{golinelli_incommensurate}. In the latter case, the
incommensurate modes originate from the fact that in the absence of
the biquadratic chain, the (gapped) modes of the spin-1 chain are at $q=0$ and
$q=\pi$ whereas at the ULS point the (gapless) modes are at $q=0$ and
$q=2\pi/3$. The presence of gapped incommensurate modes between these
two limits is merely a consequence of the continuity of the transition
between the Haldane gap phase and the gapless phase beyond the ULS
point. On the other hand, in the presence of the magnetic field, the
gapless modes of the spin-tube or those of the spin-1 chain  simply
move away from $2\pi/3$ similarly to what happens in a single
spin-1/2 chain.

\subsection{ Is there a magnetization plateau for $\langle S^z
\rangle=1$ ?}
\subsubsection{ The Umklapp terms and quantization condition on the magnetization}

\medskip

In the presence of a magnetic field, one of the central issues is the
quantization condition on the total magnetization $\langle  S^z\rangle$ for the
appearance of plateaus. This condition may be investigated by looking
at the bosonized expression for the
spin-operators (\ref{eq:lambda_operators}).  After using the transformation
(\ref{transf}) to take into account a non-zero magnetization, one can
rederive an expression for the non-SU(3) symmetric perturbations. 
Contrary to the case of zero magnetization, we cannot assume
\emph{a priori} that the ``$4k_F$'' terms are highly oscillating since
the phase $e^{4\imath k_F x}$ may be compensated by a phase arising
from the transformation (\ref{transf}). This can be interpreted as a
condition for Umklapp processes between the three different fermion flavors.
A systematic investigation indicates that the only possible Umklapp
terms   originate  from the term
$\lambda^3_i\lambda^3_{i+1}$  
or $\lambda_i^8\lambda_{i+1}^8$ in the Hamiltonian (\ref{htot}). These
Umklapp terms are:

\bea
\cos(\sqrt{8}\phi_a-2\sqrt{\frac{3}{2}} \tilde \phi_b +\alpha_1) \label{umk1}
\\
\cos(\sqrt{8}\phi_a+2\sqrt{\frac{3}{2}}\tilde \phi_b + \alpha_2 )\label{umk2}\\
\cos(4\sqrt{\frac{2}{3}}\tilde \phi_b + \alpha_3),\label{umk3}
\eea

\noindent The conditions that must be fulfilled so that these Umklapp
terms be present are:

\begin{equation}\label{eq:condition_umk1}
\langle \Lambda^3\rangle-\frac{\langle \Lambda^8\rangle}{\sqrt{3}}=\frac{1}{3} \mbox{ }\mbox{or}
\mbox{ }-\frac{2}{3},
\end{equation}
for the term (\ref{umk1}),
\begin{equation}\label{eq:condition_umk2}
\langle \Lambda^3\rangle+\frac{\langle \Lambda^8\rangle}{\sqrt{3}}=\frac{1}{3} \mbox{ }\mbox{or}
\mbox{ }-\frac{2}{3},
\end{equation}
for the term (\ref{umk2}), and
\begin{equation}\label{eq:condition_umk3}
\frac{\langle \Lambda^8\rangle}{\sqrt{3}}=-\frac{1}{6} \mbox{ }\mbox{or} \mbox{ }
\frac{1}{3},
\end{equation}
for the term (\ref{umk3}).
If we take into account the fact that $\langle \Lambda_3 \rangle=0$,
there are only two nontrivial conditions to obtain an Umklapp.
These are the condition  (\ref{eq:condition_umk2}) that reduces to
$\langle S^z \rangle = 7/6$, and the condition (\ref{eq:condition_umk3}),
 reducing to $\langle  S^z \rangle=1$. 
At these values of the magnetization the
operators (respectively) (\ref{umk2}) or (\ref{umk3}) 
  can in principle open a gap in the excitations of the system
 leading to a plateau  in the magnetization curve. In agreement with
the LSM theorem, this implies a ground state that breaks
translation invariance. 
For this to happen, these operators  must be relevant, namely,
\begin{equation}\label{eq:umk2_relevant}
(K_a+9K_b/4)<1
\end{equation}
for the term (\ref{umk2}), and 
\begin{equation}
\label{eq:umk3_relevant}
K_b<3/4
\end{equation}
for the term (\ref{umk3}).
If one of these conditions is satisfied, a gap exists 
at $\langle S^z \rangle=5/6$ more likely than at $\langle S^z
\rangle=1$ or $\langle S^z \rangle=7/6$.

\subsubsection{A theory of the possible plateau at $\langle S^z
\rangle =1$}
There seems to be evidence for an extra plateau at $\langle S^z
\rangle=1$ 
in the magnetization curve  of a 36 sites system at $J_\perp/J=3$
 obtained by DMRG in  Ref.~\onlinecite{sen} 
(see Figure~10 of Ref.~\onlinecite{sen}). 
However, this small plateau
observed at $\langle  S^z \rangle =1 $ in the  magnetization curve may
also be ascribed to the small system size\cite{diptiman_sen_pc}. Here,
we will investigate in more details the possibility of such a plateau in
the framework of the bosonized theory. In particular, we will try to give a
description of the behavior of correlation functions in the system.

Due to the presence of the Umklapp term (\ref{umk3}) the bosonized
Hamiltonian describing the low energy excitations of the spin tube at
$\langle  S^z \rangle =1 $ is:
\begin{eqnarray}
\label{eq:hamiltonian_m=1}
H&=& H_a + H_b \nonumber \\
H_a &=& \int dx \frac {dx}{2\pi} \left[ u_a K_a (\pi \Pi_a)^2 +
\frac{u_a}{K_a} (\partial_x\phi_a)^2 \right] + \frac{2g_1}{(2\pi a)^2}
\int dx \cos \sqrt{8} \phi_a \nonumber \\
H_b&=& \int dx \frac {dx}{2\pi} \left[ u_b K_b (\pi \Pi_b)^2 +
\frac{u_b}{K_b} (\partial_x\phi_b)^2 \right] + \frac{2u'a}{3 (\pi a)^2}
\int dx \cos \left( 4 \sqrt{\frac 2 3} \phi_b + \frac{2\pi} 3\right)
\end{eqnarray}
If there is indeed a magnetization plateau at $\langle  S^z
\rangle=1$, a gap opens in the excitation spectrum of the field
$\phi_b$. However, the Hamiltonian contains no terms coupling
$\phi_a$ and $\phi_b$, so that $\phi_a$ could remain ungapped. 
 This would lead to a single component 
Luttinger liquid behavior on this plateau, and a power law decay of
some spin-spin correlation functions.
Since $u' <0$, the Umklapp term would impose $\sqrt{\frac 2 3}\langle
\tilde \phi_b \rangle = -\frac{2\pi} 3$. 
If this is so, careful treatment of the expressions
(\ref{eq:lambda_operators}) of the bosonized forms of the $\Lambda$
operators is necessary  when we have a gap,
on the $\langle S^z \rangle=1$ plateau.
To eliminate gapped states completely, one can 
use the following expression (see appendix \ref{app:abelian_hubbard_su(3)}).
\begin{equation}
\Lambda^8 _i = \frac{1}{\sqrt{3}}(1 -3 c_{i,3}^\dag c_{i,3}), 
\end{equation}
for $S^z _i$ instead of (\ref{eq:abrikosov_su(3)}) here we use
the constraint (\ref{constraint}). This equation indicates
we have only the gapful excitations $\phi_b$ in $\Lambda^8(x)$. 
One has
\begin{eqnarray}
\label{eq:lambda_op_m=1}
(\Lambda^1+\imath \Lambda^2)(x)&=&\frac{e^{\imath \sqrt{2}\theta_a}}{\pi
a}\left[ 2\cos \sqrt{2} \phi_a + 2 C_1 \cos \left( \frac \pi {2a} x
+\frac \pi 6\right) \right] \nonumber \\
\Lambda^8(x)&=& \frac{3} {\pi a \sqrt{3}} e^{\imath \frac \pi a x} C_2
\end{eqnarray} 
Where:
\begin{eqnarray}
C_1=\langle e^{\imath \sqrt{ \frac 2 3} ( \tilde \phi_b - \langle
\tilde \phi_b \rangle)}\rangle. 
\nonumber \\
 C_2=\langle e^{\imath 2 \sqrt{ \frac 2 3} ( \tilde \phi_b - \langle
\tilde \phi_b \rangle)} \rangle
\end{eqnarray}
We do not give the expressions for the other operators since they 
show exponential decay 
except for $\Lambda^3$, 
never enter
spin correlation function.
We find then that translational symmetry is broken on the
$\langle S^z \rangle=1$ plateau, with a period of $4$ for the ground
state.

 This is in agreement with the LSM theorem that rules out a
translationally invariant plateau at $\langle S^z \rangle=1$.
We also see that an additional period of $4$ will appear in the
correlation functions showing a power law decay.
The correlation functions for the operators of
Eq. (\ref{eq:lambda_op_m=1}) is:
\begin{eqnarray}
&&\langle T_\tau (\Lambda^1 +\imath \Lambda^2)(x,\tau) (\Lambda^1
-\imath \Lambda^2)(x',0)\rangle  =  \frac 1 {(\pi a)^2} \left[ 2C_1^2
\left\{ (1+\frac 1 2 e^{\imath \frac{\pi x'} a}) \cos \left[\frac \pi
{2a}(x-x')\right] \right.  \right. \;\;\;\;\;\;\;\;\;\;\; \nonumber \\
&& \;\;\;\;\;\;\;\;\;\;\;-\left. \left. \frac{\sqrt{3}} 2 e^{\imath \frac {\pi x'} a} \sin
\left[\frac \pi {2a} (x-x')\right]\right\} \left(\frac {a^2} {(x-x')^2
+(u_a^* \tau)^2}\right)^{\frac 1 {2K_a}} \right. \nonumber \\   
&&\;\;\;\;\;\;\;\;\;\;\;- \left.  \left( \frac{a^2}
{(x-x')^2+(u_a^* \tau)^2}\right)^{\frac {K_a} 2 + \frac 1 {2K_a}}
\frac{(x-x')^2-(u_a^* \tau)^2}{(x-x')^2+(u_a^* \tau)^2}\right] \nonumber
\end{eqnarray}

Using  the
expressions (\ref{si}) and (\ref{siz}) of the spins in terms of
$\lambda$ matrices we find that the correlations 
$\langle S^+_n S^-_{n'}\rangle$ 
show an exponential decay whereas the correlations  $\langle S^z_{n,p} S^z_{n',p} \rangle$ follow
a power law decay. 
We have the following expressions for the equal time  spin-spin correlation
functions:
\begin{eqnarray}
\label{eq:correlation_m=1}
&&\langle S_n^z \rangle = 1- \frac{C_2}{\pi} e^{\imath \pi n}
\nonumber \\
&&\langle S_{n,p}^z S_{n',p}^z \rangle - \langle S_{n,p}^z \rangle \langle
S_{n',p}^z \rangle=\frac 2 9 
 \langle (\Lambda_n^1 +\imath
\Lambda_n^2)(\Lambda_{n'}^1 -\imath \Lambda_{n'}^2) \rangle 
\end{eqnarray}

Comparing with 
figures (6) and (7) of Ref. \onlinecite{sen}, one sees that such behavior
is not obtained in numerical calculations. This leaves two options:
one is that the system size (36 sites) in Ref. \onlinecite{sen} is too small to
observe the finite but large correlation length. This is not unreasonable,
since $K_b$ could be only slightly smaller than $3/4$. The second
possibility is that the plateau at $\langle S^z \rangle
=1$ is an artifact of the small
system size. In Sec. \ref{dmrg_m=1} it will be shown using DMRG for systems
of up to 120 sites that it is the latter possibility that is obtained. 

If we wish to obtain non-trivial plateaus  smaller
values of Luttinger parameters $K_a$ or $K_b$ are needed. 
This could be achieved by  adding a sufficiently strong
antiferromagnetic Ising
term along the chain,
\begin{equation}
\sum_{i=1} ^N \sum_{p=1} ^3 S^z _{i,p} S^z_{i+1,p}
\end{equation}
or an extra coupling, 
\begin{equation}
\sum_{i=1} ^N \sum_{p=1} ^3 \sum_{q=1} ^3\vec S_{i,p} \vec S_{i+1,q}.
\end{equation}
The extra plateaus would lie  
at $\langle S^z \rangle = 5/6, \ 1, \ 7/6$ and are allowed by an
extended LSM theorem in the case of a periodic ground state. \\

\subsubsection{Zero gap for $\langle S^z \rangle =1$ and finite size 
		scaling of DMRG results}\label{dmrg_m=1}

We continue the density matrix renormalization group\cite{white,dmrg0} 
study of the three leg ladder at $\langle S^z \rangle =1$ given by Tandon
\textit{et al.} \cite{sen}.  Their results for finite chain length show 
there might be a plateau at  $\langle S^z \rangle =1$ but the system
size is too small to draw definitive conclusions\cite{diptiman_sen_pc}. In this
section, we show that the apparent plateau at  $\langle S^z \rangle
=1$ is indeed a finite size artifact.  
In finite size study, there is a finite energy gap 
between any two nondegenerate energy levels.  We therefore use $1/N$ 
scaling to show that the energy gap scales to zero in the thermodynamic 
limit for low energy excitations with $\Delta S^{tot,z}=0$ and 
$\Delta S^{tot,z}=1$, respectively.  

The system is not dimerized since there is no gap in both 
$\Delta S^{tot,z}=0$ and $\Delta S^{tot,z}=1$ excitations.  Besides the 
information that there is a certain kind of gapless excitations 
obtained from LSM, numerical analysis gives information on whether 
$\Delta S^{tot,z}=0$ and $\Delta S^{tot,z}=1$ excitations are gapped, and 
shows whether the system is dimerized.  We conclude from both the 
DMRG calculation and the RG analysis that $\Delta S^{tot,z}=1$ 
excitations are gapless for $J_\perp$ much smaller than $J$ to 
$J_\perp$ bigger than $J$ and there is no magnetization plateau at 
$\langle S^z \rangle =1$.

We use periodic boundary condition for the rung and open boundary 
condition for the other direction, the $i$ direction in 
Eq.(\ref{ham}), in our DMRG calculation.  We calculate the lowest 
energy states for $S^z = N-2$, $N-1$, $N$, $N+1$, $N+2$ and denote 
the lowest energy in each $S^z$ sector as $E_0(S_z)$ and the second 
lowest energy as $E_1(S_z)$.  We have set $h=0$ in Eq.(\ref{ham}) 
and we use $J_\perp / J =3$  to calculate $E_0(S_z)$ and $E_1(S_z)$.  
In DMRG we keep $200$ optimized states and biggest truncation error is 
$10^{-6}$.  We  make calculations for even chain length from $N=4$ to $N=40$.

For each length $N$ we calculate the $\Delta S^{tot,z}=0$ gap by
$E_1(S_z)-E_0(S_z)$, and the $\Delta S^{tot,z}=1$ gap by 
$E_0(S_z-1)+E_0(S_z+1)-2 E_0(S_z)$.  For $m=1$, we have plot 
$E_1(S_z)-E_0(S_z)$ vis $1/N$ and $E_0(S_z-1)+E_0(S_z+1)-2 E_0(S_z)$ 
vis $1/N$ at magnetization $S_z=N$ in Fig.\ref{fig:gapscaling}.  The 
fittings to the second order of $1/N$ in the figure show that the gaps 
scale to zero in form $E_1(S_z)-E_0(S_z)\sim 1/N$ and 
$E_0(S_z-1)+E_0(S_z+1)-2 E_0(S_z) \sim 1/N$ when $N\to\infty$.  

We can see the zero gap excitations also by analyzing the spectrum.
Around $\langle S^z \rangle =1$ at length $N$ and $S_z$ we will have 
$3N/2-S_z$ doublets given in Eq.(\ref{doublets}).  When $3N/2-S_z$
is odd the ground state is doubly degenerate due to the permutation 
symmetry of the two kinds of the doublets and the two ground states
have parity $-$ respect to the $i$ to $N+1-i$ reflection symmetry.
When $3N/2-S_z$ is even, the ground state is unique, but such unique
ground states for $N$ and for $N+4$ have different reflection parity.
It suggests that there is no gap.  There is no such a parity change
from $N$ to $N+4$ for ground states of gapped translational invariant 
systems or dimerized system.  It supports the basic picture in 
previous sections: when we increase or decrease $S_z$, we put in or 
take out gapless quasiparticles (doublets here), these gapless 
quasiparticles have different parity on different energy levels. 
These parity and degeneracy can be obtained by further detailed analysis
of the low energy excitations of the Luttinger liquid.\cite{parity}

We conclude that there is no plateau at $\langle S^z \rangle =1$ since
there is no gap for $\Delta S^{tot,z}=1$ excitations.  To obtain 
numerically the non-trivial plateaus, we need smaller value of 
Luttinger parameters $K_a$ or $K_b$.  If we add a sufficiently strong 
Ising term along the chain direction (in $J$), the condition is 
satisfied and we can obtain such plateaus at 
$\langle S^z \rangle =5/6$, $1$, and $7/6$ as we predicted out in 
previous subsections.  These plateaus are 
allowed by an extended LSM theorem in the case of periodic ground state 
with a certain periodicity, as explained in section III.  
DMRG calculation for $\langle S^z \rangle =5/6$ and $7/6$ asks for 
more technique effort\cite{dmrg2}.  Calculating numerically the 
Luttinger liquid exponents\cite{dmrg3} is also important to fulfill 
the understanding of trileg ladder in future studies.

\section{Conclusions}\label{conclusions}
In this paper, we have analyzed the strong coupling limit of the three
chain system with periodic boundary conditions in the presence of a
 magnetic field, 
using a mapping on an
anisotropic SU(3) chain. A straightforward extension of the LSM
theorem allowed us to locate the possible magnetization
plateaus. Then, we applied bosonization and renormalization group
techniques to show that for $1/2<m<3/2$, the system would be described
by a two component Luttinger liquid. This allowed us to obtain the
spin-spin correlation functions of the  system, and to follow the
position of the various incommensurate modes in the spin
spin-correlation function as a function of the magnetization. Finally,
we have predicted the temperature dependence of the NMR relaxation
rate in this region.
We have also discussed the relation of the two component Luttinger
liquid we  obtained with the two component Luttinger liquid phase
of the bilinear biquadratic spin-1 chain, and shown that despite
the similarity of the two problems, the two phases are not related to
each other. 
We considered the evidence for a magnetization plateau
at $m=1$ in the framework of our bosonized description, and concluded
that if such a plateau exists, the ground state at $m=1$ should break
translational symmetry with a period of two lattice spacing. We 
obtained the correlation functions of such a ground state as well as
the average value of the magnetization at each site, and found 
that if such a plateau does exist the ground state would exhibit some
kind of antiferromagnetic order. The numerical simulation of
Ref.~\onlinecite{sen} shows no evidence for such antiferromagnetic
order thus pointing to an absence of plateau at $m=1$. 
To clarify whether or not there is a plateau at $m=1$,
we calculate energy gap around $m=1$ using DMRG method
and fit the data 
in the linear function of the inverse system size.
No gap was found in the thermodynamic limit,
therefore, we conclude that there is no plateau at $m=1$. 
One obvious direction to extend our work is to calculate numerically
the Luttinger liquid exponents of the three chain ladder with periodic
boundary conditions. In appendix~\ref{app:exposants_bosonized},
indications can
be found on how these exponents could in principle be extracted. The study 
of these exponents in the case of anisotropic
integrable SU(3) spin chains would also be interesting.
The present paper has been almost entirely concerned with the limit
$J_\perp\gg J$, and it would be interesting to investigate whether the
behavior we have obtained is valid also in the opposite limit,
$J_\perp \ll J$. It would also be interesting to study anisotropic
generalizations of the spin-tube model and check for plateaus at
$m=1$ as well as extend the analysis to higher n spin-tubes. 
Finally, the relation between the model we have considered and
classical statistical systems such as Coulomb gases may be worth
analyzing in more details.

\section{Acknowledgments}

 C.I., would like to thank I. Affleck
for kind hospitality in University of British Columbia extended to him.
E. O. acknowledges support from NSF under grant
NSF DMR 96-14999. R. Citro acknowledges financial support from
Universit\`a degli Studi di Salerno. The various authors are grateful
to I. Affleck, R. Chitra, M. Oshikawa and Paul Zinn-Justin for many useful 
and enlightening
discussions. E. O. thanks  D. Sen for e-mail correspondance on the plateaus
problem.

\vfill
\noindent $^{\star}$ {\footnotesize On leave from: Dipartimento di Scienze
Fisiche ``E. Caianiello'', Universit\`a di Salerno and Unit\`a INFM di
Salerno, 84081 Baronissi (Sa), Italy} \newline
\noindent $^{\star \star}$ {\footnotesize On leave from: Department
of Physics, Nihon University, Kanda Surugadai, Tokyo 101, Japan}
\noindent $^{\star \star \star}$ {\footnotesize On leave from: 
Institute of Theoretical Physics, CAS, Beijing 100080, P R China}

\newpage

\appendix

\section{Abelian bosonization of the SU(3) Hubbard
model}\label{app:abelian_hubbard_su(3)}
  
In this section, we apply abelian bosonization to the SU(3) Hubbard
model:
\begin{equation}
H_h=-t \sum_{i,n} \lbrack c^{\dagger}_{i,n} c_{i+1, n} + \text{h.c.}
\rbrack +\frac U 2 \sum_{i,n\ne m} n_{i,n} n_{i,m} .
\end{equation}

\noindent In the strong-coupling limit, the Hubbard Hamiltonian with
one fermion per site
projected onto the low-energy states becomes simply the Heisenberg
Hamiltonian as can be seen\cite{assaraf_su(n)} by considering
perturbation theory in 
the hopping term for $U\gg t$. Only second-order perturbation theory survives
 and the effective Heisenberg coupling is
$J=t^2/U$.

 In the continuum limit, in terms of the right and left
moving fermions introduced in Sec. \ref{weakcoupling}, the free
Hamiltonian $H_t$ can
be rewritten as:

\begin{equation}
H_t=-iv\int dx \sum_n (\psi^{\dagger }_{R,n}\partial_x \psi_{R,n}-
\psi^{\dagger }_{L,n}\partial_x \psi_{Ln}),
\end{equation}

\noindent where $v=2ta\sin(k_F a)$ is the Fermi velocity.
In the following, we are working at a filling of one fermions per
site. This implies $k_F=\pi/(3a)$ and $v_F=\sqrt{3}ta$.

 Using the
standard dictionary of abelian bosonization \cite{schulz_houches} we
express  $\psi_{R(L)n}$ in terms of the bose fields $\phi_n$ and their duals
$\theta_n$, for each flavor $n=1,2,3$,

\begin{eqnarray}
\psi_{Rn}(x)&=&\frac{1}{\sqrt{2\pi a}}e^{i(\theta_n(x)+\phi_n(x))}\eta_{Rn}
 \nonumber \\
\label{bosf}
\psi_{Ln}(x)&=&\frac{1}{\sqrt{2\pi a}}e^{i(\theta_n(x)-\phi_n(x))}\eta_{Ln},
\end{eqnarray}

\noindent where $\eta_{R(L)n}$
are the Klein factors ensuring the proper anticommutation
relations among fermion operators \cite{schulz_moriond}.
One has: $\pi\Pi_n(x)=\partial_x \theta_n$ and
$[\phi_n(x),\Pi_m(x')]=\imath \delta_{n,m} \delta(x-x')$.

The non interacting Hamiltonian is straightforwardly rewritten as:
\begin{equation}
H_t=\sum_{n=1}^3 v \int \frac{dx}{2\pi}\left[(\pi\Pi_n)^2 +(\partial_x
\phi_n)^2\right] 
\end{equation}

And the fermion densities as:
\begin{equation}\label{eq:fermion_densities}
\rho_n(x)=-\frac{\partial_x \phi_n}{\pi} + \frac{e^{-2 \imath k_F x}}{2
\pi a} e^{2 \imath \phi_n} +   \frac{e^{2 \imath k_F x}}{2
\pi a} e^{-2 \imath \phi_n}
\end{equation}
where $k_F=\frac \pi {3a}$. 

The Hubbard interaction, $V= (Ua/2) \sum_{n\ne m} \int dx \rho_n(x) \rho_m(x)$,
is rewritten in terms of the fields $\phi_n$,
\begin{equation}
V= \int dx \sum_{n\ne m} \frac{U}{\pi^2} \partial_x \phi_m \partial_x
\phi_m + \frac{2U}{(2\pi a)^2} \cos 2(\phi_n -\phi_m)  
\end{equation} 

  Instead of working with fields
$\phi_1,\phi_2,\phi_3$ it is convenient\cite{assaraf_su(n)} to introduce the
transformation:
\begin{equation}
\label{transf}
\left (
\begin{array}{l}
\phi_1 \nonumber \\
\phi_2 \nonumber \\
\phi_3 \nonumber \\
\end{array}
\right )=\left (
\begin{array}{l}
\frac{1}{\sqrt{3}} \mbox{ }\mbox{ } \frac{1}{\sqrt{2}} \mbox{ } \mbox{ }
\frac{1}{\sqrt{6}} \nonumber \\
\frac{1}{\sqrt{3}} \mbox{ } \mbox{ } \frac{-1}{\sqrt{2}} \mbox{ } \mbox{ }
\frac{1}{\sqrt{6}} \nonumber \\
\frac{1}{\sqrt{3}} \mbox{ } \mbox{ }  0 \mbox{ } \mbox{ } \frac{-2}{\sqrt{6}}
\nonumber \\
\end{array}
\right )
\left (\begin{array}{l}
\phi_c \nonumber \\
\phi_a \nonumber \\
\phi_b \nonumber \\
\end{array}\right ).
\end{equation}
and similarly for the conjugate fields. The field $\phi_c$ describes the
charge excitations, whereas the fields $\phi_{a,b}$ describe the SU(3)
spin excitations. We recover in particular the fact that the SU(3)
spin excitations are described by a conformal field theory with $C=2$
whereas the charge excitations have $C=1$.
 The  charge and spin sectors of the Hubbard Hamiltonian are then
completely separated,

\bea
H&=&H_c+H_s \\
H_c&=&\int \frac {dx}{2\pi} \left[ u_c K_c (\pi \Pi_c)^2 +\frac{u_c}{K_c}
(\partial_x \phi_c)^2 \right], \label{eq:hubbard_charge_su3}\\
H_s & = & \sum_{i=a,b}
\int\frac {dx}{2\pi} \left[ u_s K_s (\pi \Pi_i)^2 +
\frac{u_s}{K_s}(\partial_x \phi_i)^2 \right] \nonumber \\
& &\;\;+ \frac{2g}{(2\pi a)^2} \int dx \left[
\cos(\sqrt{8} \phi_a) +  
\cos\sqrt{2}(\phi_a +\sqrt{3} \phi_b)+
\cos\sqrt{2}(\phi_a -\sqrt{3} \phi_b)\right],\label{eq:hubbard_spin_su3}
\end{eqnarray}
\noindent where
\begin{eqnarray}
& & u_cK_c=v_F, \mbox{ }\mbox{ } \mbox{ } \frac{u_c}{K_c}
=v_F+\frac{2Ua}{\pi} \nonumber \\
& &  u_sK_s=v_F, \mbox{ }\mbox{ } \mbox{ } \frac{u_s}{K_s}
=v_F-\frac{Ua}{\pi}, \mbox{ } \mbox{ } \mbox{ } g=Ua 
\end{eqnarray}
Note that the Hamiltonian describing the charge modes contains
\emph{no} Umklapp term that could lead to a gap opening. This is due
to the fact that Eq. (\ref{eq:fermion_densities}) has been truncated
at the $2k_F$ harmonics. In a more complete expression
\cite{haldane_bosons}, higher harmonics would appear and would give a
higher order $6k_F$-Umklapp. This Umklapp term can also be derived by
perturbation theory\cite{assaraf_su(n)}.
This Umklapp term is of the form $\cos
2\sqrt{3} \phi_c$ and is \emph{irrelevant} for $U/t \ll 1$. 
This is confirmed by  numerical simulations\cite{assaraf_su(n)} which
show  that the charge gap in a
SU(3) Hubbard model open only for $U> 2.2 t$. When considering RG 
equations (see Sec. \ref{weakcoupling})
 we shall see that in the spin
sector, for $U$ initially positive, $K_s$ renormalizes to $1$ and $g$
renormalizes to $0$. This implies that the spin sector of the SU(3)
Hubbard is described by a $c=2$ conformal field theory perturbed by a
marginally irrelevant operator\cite{aff_wz,assaraf_su(n)}.

Of course, we also need a bosonized expression for the SU(3) spin
operators. This can be derived from the continuum limit
of the definition (\ref{eq:abrikosov_su(3)}) of these operators, recall:
$\Lambda^{\alpha}(x) \simeq a^{-1}\Lambda_i^{\alpha},\;\; x=ia \;\;\;$ ($\alpha=1 \cdots 8$).
We obtain:
\begin{eqnarray}\label{eq:lambda_operators}
\Lambda^1(x)& = & \frac{\cos \sqrt{2} \theta_a }{\pi a}\left[ 2 \cos
\sqrt{2} \phi_a + e^{\imath \frac{ 2\pi}{3a}x} e^{-2 \imath
\frac{\phi_b}{\sqrt{6}}} +e^{-\imath \frac{ 2\pi}{3a}x} e^{2 \imath
\frac{\phi_b}{\sqrt{6}}} \right] \nonumber \\
\Lambda^2(x)& = & \frac{\sin \sqrt{2} \theta_a }{\pi a}\left[ 2 \cos
\sqrt{2} \phi_a + e^{\imath \frac{ 2\pi}{3a}x} e^{-2 \imath
\frac{\phi_b}{\sqrt{6}}} +e^{-\imath \frac{ 2\pi}{3a}x} e^{2 \imath
\frac{\phi_b}{\sqrt{6}}} \right] \nonumber \\
\Lambda^3(x)& = &-\frac{\sqrt{2}}{\pi}\partial_x \phi_a +
\left[\frac{\imath}{\pi
a}e^{\imath \frac{ 2\pi}{3a}x} e^{-2 \imath
\frac{\phi_b}{\sqrt{6}}} \sin \sqrt{2} \phi_a +\text{h.c.} \right]
\nonumber \\
\Lambda^4(x)& = &\frac 1 {\pi a}\cos\left(\frac{\theta_a}{\sqrt{2}}
+\sqrt{\frac 3 2} \theta_b\right)\left[ 2 
\cos\left(\frac{\phi_a}{\sqrt{2}}+\sqrt{\frac 3 2} \phi_b\right) +
e^{\imath \frac{ 2\pi}{3a}x} e^{\imath
\left(\frac{\phi_a}{\sqrt{2}}-\frac{\phi_b}{\sqrt{6}}\right)}
+\text{h.c.}\right] \nonumber \\
\Lambda^5(x) & = & \frac 1 {\pi a}\sin \left(\frac{\theta_a}{\sqrt{2}}
+\sqrt{\frac 3 2} \theta_b\right)\left[ 2 
\cos\left(\frac{\phi_a}{\sqrt{2}}+\sqrt{\frac 3 2} \phi_b\right) +
e^{\imath \frac{ 2\pi}{3a}x} e^{\imath
\left(\frac{\phi_a}{\sqrt{2}}-\frac{\phi_b}{\sqrt{6}}\right)}
+\text{h.c.}\right] \nonumber \\
\Lambda^6(x) & = & \frac 1 {\pi a}\cos\left(\frac{\theta_a}{\sqrt{2}}
-\sqrt{\frac 3 2} \theta_b\right)\left[ 2 
\cos\left(\frac{\phi_a}{\sqrt{2}}-\sqrt{\frac 3 2} \phi_b\right) +
e^{\imath \frac{ 2\pi}{3a}x} e^{\imath
\left(\frac{\phi_a}{\sqrt{2}}+\frac{\phi_b}{\sqrt{6}}\right)}
+\text{h.c.}\right] \nonumber \\
\Lambda^7(x) & = & \frac 1 {\pi a}\sin\left(\frac{\theta_a}{\sqrt{2}}
-\sqrt{\frac 3 2} \theta_b\right)\left[ 2 
\cos\left(\frac{\phi_a}{\sqrt{2}}-\sqrt{\frac 3 2} \phi_b\right) +
e^{\imath \frac{ 2\pi}{3a}x} e^{\imath
\left(\frac{\phi_a}{\sqrt{2}}+\frac{\phi_b}{\sqrt{6}}\right)}
+\text{h.c.}\right] \nonumber \\
\Lambda^8(x)& = &-\frac{\sqrt{2}}{\pi}\partial_x \phi_b
+\frac{e^{\imath \frac{ 2\pi}{3a}x}}{\pi \sqrt{3} a}\left[ e^{- \imath
\sqrt{\frac 2 3}\phi_b} \cos \sqrt{2} \phi_a -e^{ \imath
\sqrt{\frac 8 3}\phi_b} \right] +\text{h . c.}
\end{eqnarray}
Where $\Lambda^{\alpha}(x)=\frac {\Lambda^{\alpha}_n} a$ for $x=na$.
Using these expressions, one can derive immediately the
expressions (\ref{h11})--(\ref{h3}).
In the limit $U \to \infty$, one must note that the expression of
$\Lambda_8(x)$ has to be modified. The reason is the following: 
for $U\to \infty$, one has $c_1^\dagger c_1 + c_2^\dagger c_2 +
c_3^\dagger c_3=1$ on each site. As a result, $\lambda_8= (c_1^\dagger
c_1 + c_2^\dagger c_2 -2 c_3^\dagger c_3)/\sqrt{3}$ can be rewritten
as: $\lambda_8=(1-3 c_3^\dagger c_3)/\sqrt{3}$. 
Using bosonized expressions, one obtains:
\begin{equation}
\label{eq:lambda8_u=inf}
\Lambda_8(x)=-\frac{\sqrt{2}}{\pi}\partial_x \phi_b -
\frac{\sqrt{3}}{2\pi a}\left[ e^{\imath \frac {2\pi}{3a} x}
e^{-\imath\sqrt{\frac 8 3}\phi_b} +\text{H. c.}\right]
\end{equation}
Thus, the terms containing $\cos \sqrt{2} \phi_a$ drops from the
expression of $\Lambda_8(x)$ in the limit $U\to \infty$. This means
that the SU(3) Hubbard model with a finite charge gap and the SU(3)
spin chain should have in general different correlations for
$\Lambda_8$. It should be noted that this difference should not appear
at the isotropic point, where the exponents of the correlation
functions are identical. However, it is obtained for models in which
the SU(3) rotation symmetry is broken.

\section{Operator product expansion of marginal
operators}\label{app:derivation_ope} 

In this section, our aim is to derive the OPE for operators of the
form $\cos (\sqrt{8} \vec \alpha . \vec \phi)$, and deduce the
renormalization group equations.

Let us first recall briefly how operator product expansions can be used to
obtain one loop renormalization group expansions\cite{cardy_scaling}. 
Assume  we are given a set of operators $\Phi_k$,  closed under the OPE 
\bea \label{eq:ope_general}
\Phi_i(x,\tau) \Phi_j(0) \sim \sum_k  {c_{ij}}^k(x,\tau) \Phi_k(0),
\eea
in the sense that  expression (\ref{eq:ope_general}), 
when inserted in any correlation
function, gives the correct leading asymptotics for $(x,\tau) \to 0$. 
Denoting $[\Phi_i]$ the scaling dimension of the operator $\Phi_i$,
defined by :
\begin{equation} 
\langle \Phi_i(x,0) \Phi_i(0,0)\rangle \propto \left(\frac1 x  \right)^{2[\Phi_i]}
\end{equation}  
then $c_{ijk}(x,0)\sim \mathrm{const.} x^{[\Phi_k]-[\Phi_i]-[\Phi_j]}$.

Perturbing the Hamiltonian by 
\bea
H_I= \sum_k \int dx d\tau g_k \Phi_k(x,\tau)
\eea
One can deduce the one loop
 renormalization group beta functions  directly
from the OPE (\ref{eq:ope_general}). These one loop renormalization
group equations are:
\bea
 \frac{d g_k}{d l} \equiv \beta_k({ g}) = (2-[\Phi_k]) g_k -\pi \sum_{i,j} {C_{ij}}^k g_i g_j
\label{rg1}
\eea

Where we define
\begin{equation}
C_{ij}^k=a^2\int_0^{2\pi} \frac{d\theta}{2\pi} c_{ijk}(a\cos
\theta,\frac a u \sin \theta)
\end{equation}
The Equations \ref{rg1} are a slight generalizations of those that can
be found in Ref.~\onlinecite{cardy_scaling}, in which  we have allowed
for a function $c_{ijk}(x,\tau)$ that depends both on $x^2+u^2\tau^2$
and $u\tau/x$. 
The set of operators $\Phi_k$ has to be closed under the operator
product expansion (i.e. they have to form a closed algebra), 
otherwise new operators would be generated under
RG, and new OPEs would have to be derived. We thus need to retain the
smallest closed algebra that contains all the operators that appear in
our problem. 
In our case, we have to retain the operators $\cos \sqrt{8} \vec
\alpha_i.\vec \phi$ as well as the operators $(\partial_x \phi_{a,b})^2$. 

In order to derive the OPE for the $\cos \sqrt{8} \vec
\alpha_i.\vec \phi$ operators,we use the following identity:
\begin{equation}
e^{\imath \vec \alpha . \vec \phi}=e^{-\frac{\langle (\vec \alpha
. \vec \phi)^2\rangle}{2}} :e^{\imath \vec \alpha . \vec \phi}:
\end{equation}
Where $:\ldots:$ represents normal ordering.
This identity implies:
\begin{equation}
e^{\imath \sqrt{8} \vec \alpha . \vec \phi(x,\tau)}e^{-\imath \sqrt{8}\vec \alpha
. \vec \phi(0,0)} =e^{-4\langle (\vec \alpha . (\vec
\phi(x,\tau)-\vec \phi(0,0)))^2 \rangle}:e^{\imath \sqrt{8} \vec \alpha .
( \vec \phi(x,\tau) - \vec \phi(0,0))}:
\end{equation}

We have: 
\begin{equation}
e^{-4\langle (\vec \alpha . (\vec
\phi(x,\tau)-\vec \phi(0,0)))^2
\rangle}=\left(\frac{a^2}{x^2+(u\tau)^2}\right)^2K 
\end{equation}

And we can expand the normal ordered product, yielding:
\begin{equation}
:e^{\imath \sqrt{8} \vec \alpha .
( \vec \phi(x,\tau) - \vec \phi(0,0))}:=1- 4 (\vec \alpha (x \partial_x
\vec \phi(0,0) +\tau \partial_\tau \vec \phi(0,0)))^2
\end{equation}

This leads to the OPE (\ref{opem}).

Now consider:
\begin{equation}
e^{\imath \sqrt{8} \vec \alpha . \vec \phi(x,\tau)}e^{\imath \sqrt{8}\vec \beta
. \vec \phi(0,0)} =e^{-4\langle (\vec \alpha . \vec
\phi(x,\tau) + \vec \beta . \vec \phi(0,0)))^2 \rangle}:e^{\imath
\sqrt{8}( \vec \alpha . \vec \phi(x,\tau) + \vec \beta . \vec \phi(0,0))}:
\end{equation}
Rewriting it
\begin{equation}
\langle (\vec \alpha . \vec
\phi(x,\tau) + \vec \beta . \vec \phi(0,0)))^2 \rangle=\langle((\vec \alpha +
\vec \beta).\vec \phi)^2\rangle -2 \vec \alpha \vec \beta (\langle
\phi^2(0,0) - \phi(x,\tau)\phi(0,0)\rangle),
\end{equation}
we have
\begin{equation}
e^{\imath \sqrt{8} \vec \alpha . \vec \phi(x,\tau)}e^{\imath \sqrt{8}\vec \beta
. \vec \phi(0,0)} = \left( \frac a {\sqrt{x^2+u^2 \tau^2}}\right)^{-2K
\vec \alpha.\vec \beta} e^{\imath \sqrt{8} (\vec \alpha +\vec
\beta).\vec \phi}
\end{equation}
The OPE for the cosines can be obtained trivially. 
There is no need to calculate the OPEs of the operators $(\partial_x
\phi_{a,b})^2$ with the operators  $\cos \sqrt{8} \vec
\alpha_i.\vec \phi$. It can be shown that these OPEs only reflect the
dependence of the scaling dimensions of the cosines with
$K_{a,b}$. Therefore, we have obtained all the  OPEs needed to derive
the renormalization group equations.
 It is then a simple matter to write
the renormalization group equation using the formula \ref{rg1}.  
Following the procedure described
in Ref. \cite{cardy_scaling}, one obtains:                                     
\begin{eqnarray}\label{eq:rge_noexpand} 
\frac {dy_1}{dl}&=&(2-2K_a) y_1 - \frac{y_2 y_3}{2\pi^2} \nonumber \\
\frac{dy_2}{dl}&=&\left(2-\frac {K_a} 2 -3 \frac{K_b}2 \right)y_2
-\frac{y_1 y_3} 2 \nonumber \\
\frac{dy_3}{dl}&=&\left(2-\frac {K_a} 2 -3 \frac{K_b}2 \right)y_3
-\frac{y_1 y_2} 2 \nonumber \\
\frac{d}{dl}\left(  \frac 1 {K_a} \right)&=&\frac {y_1^2} 2 + \frac
{y_2^2} 8 + \frac{y_3^2} 8 \nonumber \\
\frac{d}{dl}\left(  \frac 1 {K_b} \right)&=&3 \frac
{y_2^2} 8 + 3 \frac{y_3^2} 8 
\end{eqnarray} 
where $y_i=\frac {g_i}{\pi v_F}$. A few remarks on these equations
have to be made. In Ref. \onlinecite{cardy_scaling}, the OPE depends
only on the distance between points. In our case, the OPE also depend
on the angle between the segment joining the points and the horizontal
axis. Since in the derivation of the RG equations one integrates over
the ring $a < r <ae^{dl}$ the angular part of the integration cancels
the terms $\partial_x \phi \partial_\tau \phi$ and gives a $\pi/2$
factor for the terms $(\partial_{x,\tau} \phi)^2$. The second
important remark is that in our equations, we are working with
$y_2(0)=y_3(0)$. It can be checked that this condition is preserved by
the RG flow and that under such condition no terms $\partial_x \phi_a
\partial_x \phi_b$ are generated.
Finally, if we expand for small $y_4,y_5$ , we have $K_a=1-y_4$ and
$K_b=1-y_5$. Putting this in Eqs. (\ref{eq:rge_noexpand}) we get the
renormalization group equations 

\section{Renormalization group equations in presence of the external
magnetic field}\label{rghh}
In the present section, we want to extend the derivation of the
renormalization group equations of appendix \ref{app:derivation_ope}
to the case of a non zero effective magnetic field. 
As explained in section \ref{correlations}, it is convenient to first
perform a Legendre transformation and work at fixed magnetization. 
Then, the RG equation for the magnetization becomes trivial, but the
RG equation for the magnetic field is not. 
Similarly to the zero magnetic field case, the renormalization group
equations can be obtained via operator product expansion. The only
difficulty is that the problem is not \emph{a priori} translationally
invariant. 

The relevant operator product expansions are obtained by the method of
appendix \ref{app:derivation_ope}. One has:
\begin{eqnarray}
&&\cos \left(\sqrt{8}\vec \alpha . (\vec \phi(x,\tau)+\vec u
x) \right) \cos \left(\sqrt{8} \vec \alpha . (\vec \phi(x',\tau)+\vec u
x')\right) = \frac {a^4}{(x^2+u^2 \tau^2)^2} \nonumber \\ 
&&\times\left( -\sqrt{2} \sin
\left( \sqrt{8} (\vec u . \vec \alpha) x \right) (x \partial_x (\vec
\alpha . \vec \phi) - \tau \partial_\tau (\vec \alpha . \vec \phi)) -2
\left[ (\vec \alpha .( x\partial_x \vec \phi + \tau \partial_\tau \vec
\phi)) \right]^2 \cos \left(\sqrt{8}(\vec u .\vec \alpha)x \right) \right)
\end{eqnarray}

In our case, one must take $\vec u=-\pi m_b (0,1)$. The term in $\tau
\partial_\tau (\vec \alpha. \vec \phi)$ disappears upon angular
integration. On the other hand, the term in $ x \partial_x (\vec
\alpha. \vec \phi)$ leads to a renormalization of the applied magnetic
field. The angular integrations lead in general to Bessel Functions. 

The second useful OPE is:
\bea
e^{\imath \sqrt{8} \vec \alpha . (\vec \phi(x,\tau) +\vec u x)}e^{\imath \sqrt{8}\vec \beta
. (\vec \phi(x',0)+\vec u x')}& = & \left( \frac a {\sqrt{(x-x')^2+u^2
(\tau-\tau')^2}}\right)^{-2K
\vec \alpha.\vec \beta} e^{\imath \sqrt{8} (\vec \alpha +\vec
\beta).(\vec \phi(\frac {x+x'} 2,0)+ \vec u \frac {x+x'} 2) }\nonumber \\
& \times & e^{\imath (\vec \alpha -\vec \beta).\vec u \frac{x-x'} 2} 
\eea

These OPEs allow us to deduce the renormalization group equations for
$K_a,K_b,y_1,y_2,h$ in the form:
\begin{eqnarray}
\frac d {dl} \left(\frac 1 {K_a} \right)&= &\frac 1 2 y_1^2 +\frac 1 4
y_2^2 J_0(\pi m_b(l) \frac{\sqrt{3}}{2} a) \nonumber \\
\frac d {dl} \left(\frac 1 {K_b}\right)&=& \frac 3 4
y_2^2 J_0(\pi m_b \sqrt{\frac{3}{2}} a(l)) \nonumber \\
\frac {dy_1}{dl}&=&(2-2K_a) y_1 -\frac 1 2 y_2^2 J_0(\pi m_b(l)
\sqrt{3}a) \nonumber \\
\frac{dy_2}{dl}&=&\left(2-\frac 1 2 K_a -\frac 3 2 K_b\right) y_2 - y_1
y_2 J_0(\pi \frac{\sqrt{3}} 2 m_b(l) a) \nonumber \\
\frac{dh}{dl}&=& \sqrt{\frac 3 {8a}} y_2^2 J_1(\pi \sqrt{6} m_b(l) a)  
\end{eqnarray} 
 
\section{Determination of the exponents of the bosonized
Hamiltonian}\label{app:exposants_bosonized}
In this section, we discuss the determination of the exponents of the
spin-tube. We have shown previously that in weak coupling the model
flows to a two-component Luttinger liquid fixed point. In strong
coupling some alternative 
techniques are needed to determine the Luttinger liquid
exponents from thermodynamic quantities. Note that in the isotropic SU(N)
Hubbard model case\cite{assaraf_su(n)} with a charge gap, 
one needs only the spin
velocity since the spin exponents are constrained by SU(N)
invariance. Here, SU(3) symmetry is broken and we expect two different
velocities$u_a,u_b$ and two exponents $K_a,K_b$ so that we need four
independant quantities.

Suppose that we have a two component Luttinger liquid described by the
Hamiltonian: 
\begin{equation}
H=\sum_{i=a,b}\int \frac{dx}{2\pi} \left[u_i K_i (\pi \Pi_i)^2
+\frac{u_i}{K_i} (\partial_x \phi_i)^2 \right]
\end{equation}
In the case of the spin tube, we can rule out  terms of the form
$\Pi_a \Pi_b$ and $\partial_x \phi_a \partial_x \phi_b$ since we know
that they are not present in the bare Hamiltonian and not generated
upon RG. Moreover, we know that $\chi_{38}=0$ which guarantees the
absence of terms $\partial_x \phi_a \partial_x \phi_b$ in the
Hamiltonian.   

The usual technique to determine the Luttinger liquid exponents is to
consider the energy change induced by taking $\pi \langle \Pi_{a,b}\rangle=
\varphi_{a,b}/L$. 
In terms of the Luttinger liquid parameters, this energy change is
given by:
\begin{equation}
\delta E = \frac{u_a K_a}{2\pi L} (\varphi_a)^2 + \frac{u_b K_b}{2 \pi L} (\varphi_b)^2
\end{equation}  
This energy change is related to the change of ground state energy
caused by taking twisted boundary conditions.
Let us discuss in more details these twisted boundary conditions in
the specific case of the spin tube.
 Using the bosonization formulas, one sees easily that:
\begin{eqnarray}
\Lambda^1& + & \imath \Lambda^2 \propto e^{\imath \sqrt{2}\theta_a}
\nonumber \\ 
\Lambda^4& + & \imath \Lambda^5 \propto e^{\imath
\left(\frac{\theta_a}{\sqrt{2}}+\sqrt{\frac 2
3}\theta_b\right)}\nonumber \\
\Lambda^6& + & \imath \Lambda^7 \propto e^{\imath
\left(-\frac{\theta_a}{\sqrt{2}}+\sqrt{\frac 2
3}\theta_b\right)}
\end{eqnarray}

Therefore, imposing $\langle \pi \Pi_a \rangle =\varphi_a/L$ and
$\langle \pi \Pi_b \rangle = \varphi_b/L$ amounts to imposing the
boundary conditions:
\begin{eqnarray}\label{eq:twisted_bc_tube}
(\Lambda^1+ \imath \Lambda^2)(L)& = & (\Lambda^1+\imath
\Lambda^2)(0)e^{\imath \sqrt{2} \varphi_a}\nonumber
\\
(\Lambda^4+  \imath \Lambda^5)(L)& = & (\Lambda^4+\imath
\Lambda^5)(0)e^{\imath (\varphi_a/\sqrt{2} + \sqrt{\frac 2
3} \varphi_b) }\nonumber \\
(\Lambda^6 +  \imath \Lambda^7)(L)& = & (\Lambda^6+\imath
\Lambda^7)(0)e^{\imath(\sqrt{\frac 2
3} \varphi_b -\varphi_a/\sqrt{2})}
\end{eqnarray}
As an aside, one should remark that the transformation:$\Pi(x) \to
\Pi(x)-f(x)$ is realized by the operator $U=\exp \left( -\imath \int dx
f(x) \phi(x) \right)$. This operator can also be written as:
$U=\exp \left( \imath \int dx
F(x) \partial_x \phi(x) \right)$ where $f=\frac {dF}{dx}$.
Twisted boundary conditions correspond to $f(x)=\alpha/L$. Therefore,
an operator generating states satisfying boundary conditions
(\ref{eq:twisted_bc_tube}) acting on states satisfying periodic
boundary conditions can be built in the continuum. A lattice version
is easily constructed,  giving an operator  of the form:
\begin{eqnarray}
U(\varphi)=\exp\left(-\imath \sum_{n=1}^L \frac{(n-1)}{L} (\varphi_a
\Lambda_n^3 +\varphi_b
\Lambda_n^8)\right) 
\end{eqnarray}
 One can check that these lattice  operators acting on states that
satisfy periodic boundary conditions
 generate states that satisfy the boundary conditions
(\ref{eq:twisted_bc_tube}) directly on the
lattice. This guarantees the existence of states satisfying the
twisted boundary conditions (\ref{eq:twisted_bc_tube}). 
The generalization of this construction to the SU(N) case is
trivial. Instead of $\Lambda^{3,8}$ one has to consider the Maximal
Abelian Subalgebra (MASA)  and builds the operators
corresponding to $U(\varphi)$.
In the case of the spin-tube, the energy change of the ground state
obeys the condition:
\begin{equation}
L \delta E(\varphi_a, \varphi_b)=L \delta E(\varphi_a,0) +L \delta E(0, \varphi_b) +o(\varphi_a^2,\varphi_b^2)
\end{equation} 

In order to determine completely $u_a,u_b, K_a, K_b$ 
suppose that one places the anisotropic SU(3) chain in fields that
couple to the $\Lambda^3$ and $\Lambda^8$ components of the spin,
\begin{equation}
H=\sum_i \sum_{\alpha} a_\alpha \lambda_i^\alpha \lambda_{i+1}^\alpha
-h_3 \sum_i \lambda_i^3 -h_8  \sum_i \lambda_i^8
\end{equation}
Then, the Hamiltonian becomes in the continuum:
\begin{equation}
H=H_0+\sum_{\nu=a,b}h_\nu \frac{\partial_x\phi_\nu}{\pi}
\end{equation}
where $h_a=h_3$, $h_b=h_8$.
Then, one has:
\begin{equation}
 \frac{-\langle \partial_x \phi_\nu \rangle }{\pi}
=\frac {K_\nu}{u_\nu} h_\nu
\end{equation}
Thus, one has:
\begin{eqnarray}
\langle  \Lambda^3 \rangle & = & K_a/u_a h_3 \nonumber \\
\langle  \Lambda^8 \rangle & = & K_b/u_b h_8
\end{eqnarray}

This is sufficient  to extract the parameters of the
two component Luttinger liquid associated with the anisotropic SU(3)
spin chain. However, we started from  three coupled spin 1/2  chains with
periodic boundary conditions. To extract the two-component
Luttinger liquid exponents for this problem, we need to express the
preceding formulas in terms of the original spins. 
Re-expressing the twisted boundary conditions in terms of the original
spins is elementary if one remembers that when we choose:
$S^z=(5/6-\lambda^8/\sqrt{3})$, we have:
\begin{eqnarray}
(\lambda^6-\imath \lambda^7)_i & = &\frac 2 {\sqrt{3}} \sum_p j^{p-1}
S_{i,p}^+ \nonumber \\
(\lambda^4-\imath \lambda^5)_i & = &\frac 2 {\sqrt{3}} \sum_p j^{2(p-1)}
S_{i,p}^+ \nonumber \\
(\lambda^1-\imath \lambda^2)_i& = & - \frac 1 2 \sum_p j^{2(p-1)}
S_{i,p}^z 
\end{eqnarray}
The following expression for $\lambda^3$ can also be obtained:
\begin{equation}
\lambda_3=\frac{\imath}{\sqrt{3}} \left((S_2^-S_1^+ -S_2^-S_1^+) +
(S_3^-S_2^+ - S_3^+S_2^-) + (S_1^-S_3^+ - S_1^+S_3^-)\right) P_{S^z=1/2},
\end{equation}
where $P_{S^z=1/2}$ is the projector on the subspace $S^z=1/2$.
Physically, $\lambda_3$ is proportional to the spin current in the
transverse direction. It is therefore $+1$ for positive chirality and
$-1$ for negative chirality. It can also be rewritten as:
\begin{equation}
\lambda_3=\frac 2 {\sqrt{3}} \left[ (\vec S_2 \times \vec S_1)^z
+(\vec S_3 \times \vec S_2)^z + (\vec S_1 \times \vec S_3)^z \right]
P_{S^z=1/2}
\end{equation}
With these expressions, it is in principle possible to obtain
numerically the Luttinger liquid exponents for a general three leg spin ladder
with periodic boundary conditions under magnetic field between the
$\langle S^z \rangle =1/2$ and $\langle S^z\rangle =3/2$ plateaus.

\bigskip


\begin{thebibliography}{10}

\bibitem{ladder}
E. Dagotto and T. Rice, Science {\bf 271},  618  (1996).

\bibitem{haldane_gap}
F.~D.~M. Haldane, Phys. Rev. Lett. {\bf 50},  1153  (1983).

\bibitem{azuma_srcuo}
M. Azuma {\it et~al.}, Phys. Rev. Lett. {\bf 73},  3463  (1994).

\bibitem{chaboussant_cuhpcl}
G. Chaboussant {\it et~al.}, Phys. Rev. B {\bf 55},  3046  (1997).

\bibitem{schulz_berlin}
H. Schulz,  in {\em Strongly Correlated Magnetic and Superconducting Systems},
  Vol.~478 of {\em Lecture Notes in Physics}, edited by G. Sierra and M.
  Martin-Delgado (Springer, Berlin, 1997).
\bibitem{orignac_spintube}
E. Orignac, R. Citro, and N. Andrei, Low-energy behavior of the spin-tube model
  and coupled XXZ chains, 1999, article in preparation.

\bibitem{oshikawa}
M. Oshikawa, M. Yamanaka, and I. Affleck, Phys. Rev. Lett. {\bf 78},  1984
  (1997).

\bibitem{cabra}
D. Cabra, A. Honecker, and P. Pujol, Phys. Rev. Lett. {\bf 79},  5126  (1997).

\bibitem{cabra_unp}
D. Cabra, A. Honecker, and P. Pujol, Phys. Rev. B {\bf 58},  6241  (1998),
  cond-mat/9802035.

\bibitem{cithra}
R. Chitra and T. Giamarchi, Phys. Rev. B {\bf 55},  5816  (1997).

\bibitem{affleck_houches}
I. Affleck,  in {\em Fields, Strings and Critical Phenomena}, edited by E.
  Brezin and J. Zinn-Justin (Elsevier Science Publishers, Amsterdam, 1988).

\bibitem{reigrotzki}
M. Reigrotzki, H. Tsunetsugu, and T.~M. Rice, J. Phys. Condens. Matter {\bf 6},
   9235  (1994).

\bibitem{white}
S.~R. White, Phys. Rev. Lett. {\bf 69},  2863  (1992).

\bibitem{dmrg0}
S.~R. White, Phys. Rev. B {\bf 48},  10345  (1993).

\bibitem{kawan1}
K. Kawano and M. Takahashi, J. Phys. Soc. Jpn. {\bf 66},  4001  (1997).

\bibitem{mila_field}
F. Mila, Eur. Phys. J. B {\bf 6},  201  (1998), cond-mat/9805029.

\bibitem{lsm}
E. Lieb, T.~D. Schultz, and D. Mattis, Ann. Phys. {\bf 16},  407  (1961).

\bibitem{sen}
K. Tandon {\it et~al.}, Phys. Rev. B {\bf 59},  396  (1999), cond-mat/9806111.

\bibitem{prokovskii}
S. Pokrovskii and A.~M. Tsvelik, Sov. Phys. JETP {\bf 66},  1275  (1987).

\bibitem{fath_biquadratic}
G. F\'ath and P.~B. Littlewood, Phys. Rev. B {\bf 58},  R14709  (1998),
  cond-mat/9810259.

\bibitem{gellmann_matrix}
E. Gusev, Theor. Math. Phys. {\bf 53},  1018  (1983).

\bibitem{gasiorowicz}
S. Gasiorowicz, {\em Elementary Particle Physics} (Wiley, New York, 1966), p.
  261.

\bibitem{uimin}
G. Uimin, JETP Lett. {\bf 12},  225  (1970).

\bibitem{sutherland}
B. Sutherland, Phys. Rev. B {\bf 12},  3795  (1975).

\bibitem{parkinson_uls_magfield}
J.~B. Parkinson, J. Phys. Condens. Matter {\bf 1},  6709  (1989).

\bibitem{kiwata_uls_magfield}
H. Kiwata, J. Phys. Condens. Matter {\bf 7},  7991  (1995).

\bibitem{schmitt_uls_phase_diagram}
A. Schmitt, K.~H. M\"utter, and M. Karbach, J. Phys. A {\bf 29},  3951  (1996).

\bibitem{maassarani_su(n)_xx}
Z. Maassarani and P. Mathieu, Nucl. Phys. B {\bf 517},  395  (1998),
  cond-mat/9709163.

\bibitem{lsm_theorem}
E. Lieb, T.~D. Schultz, and D. Mattis, Ann. Phys. {\bf 16},  407  (1961).

\bibitem{affleck_lieb}
I. Affleck and E.~H. Lieb, Lett. Math. Phys. {\bf 12},  57  (1986).

\bibitem{aff_wz}
I. Affleck, Nucl. Phys. B {\bf 265},  409  (1986).

\bibitem{affleck_su(n)}
I. Affleck, Nucl. Phys. B {\bf 305},  529  (1988).

\bibitem{witten_wz}
E. Witten, Commun. Math. Phys. {\bf 92},  455  (1984).

\bibitem{tsvelikb}
A. Tsvelik, {\em Quantum Field Theory in Condensed Matter Physics} (Cambridge
  University Press, Cambridge, 1995).

\bibitem{affleck_log_corr}
I. Affleck, D. Gepner, T. Ziman, and H.~J. Schulz, J. Phys. A {\bf 22},  511
  (1989).

\bibitem{ziman_spin3/2}
T. Ziman and H.~J. Schulz, Phys. Rev. Lett. {\bf 59},  140  (1987).

\bibitem{schulz_hubbard_exact}
H.~J. Schulz, Phys. Rev. Lett. {\bf 64},  2831  (1990).

\bibitem{schulz_houches}
H.~J. Schulz,  in {\em Proceedings of Les Houches Summer School LXI}, edited by
  E. Akkermans, G. Montambaux, J. Pichard, and J. Zinn-Justin (North-Holland,
  Amsterdam, 1995), p.\ 533.

\bibitem{kawakami_bethe_U<0}
N. Kawakami and S.~K. Yang, Phys. Rev. B {\bf 44},  7844  (1991).

\bibitem{hayward_2chain}
C.~A. Hayward {\it et~al.}, Phys. Rev. Lett. {\bf 75},  926  (1995).

\bibitem{mila_zotos}
F. Mila and X. Zotos, Europhys. Lett. {\bf 24},  133  (1993).

\bibitem{ogata_tj}
M. Ogata, M.~U. Luchini, S. Sorella, and F.~F. Assaad, Phys. Rev. Lett. {\bf
  66},  2388  (1991).

\bibitem{assaraf_su(n)}
R. Assaraf, P. Azaria, M. Caffarel, and P. Lecheminant, Metal-insulator
  transition in the one-dimensional SU(N) Hubbard model, 1999,
  cond-mat/9903057.

\bibitem{itoikato}
C. Itoi and M.-H. Kato, Phys. Rev. B {\bf 55},  8295  (1997).

\bibitem{itoimukaida}
C. Itoi and H. Mukaida, J. Phys. A {\bf 27},  4695  (1994).

\bibitem{note_fusion_2d}
It is interesting to note that a similar Hamiltonian enters in the theory of
  two dimensional melting (see eq. (3.16) in Ref.
  \onlinecite{nelson_fusion_2d}).

\bibitem{kosterlitz_renormalisation_xy}
J.~M. Kosterlitz, J. Phys. C {\bf 7},  1046  (1974).

\bibitem{jose}
J. Jos\'e, L. Kadanoff, S. Kirkpatrick, and D. Nelson, Phys. Rev. B {\bf 16},
  1217  (1977).

\bibitem{giamarchi_logs}
T. Giamarchi and H.~J. Schulz, Phys. Rev. B {\bf 39},  4620  (1989).

\bibitem{nelson_fusion_2d}
D.~R. Nelson, Phys. Rev. B {\bf 18},  2318  (1978).

\bibitem{kadanoff_gaussian_model}
L.~P. Kadanoff, Ann. Phys. {\bf 120},  39  (1979).

\bibitem{knops_sine-gordon}
H.~J.~F. Knops and L.~W.~J. {Den Ouden}, Physica A {\bf 103},  597  (1980).

\bibitem{itoi_rg_calculation}
C. Itoi (unpublished).

\bibitem{haldane_xxzchain}
F.~D.~M. Haldane, Phys. Rev. Lett. {\bf 45},  1358  (1980).

\bibitem{giamarchi_spin_flop}
T. Giamarchi and H.~J. Schulz, J. Phys. (Paris) {\bf 49},  819  (1988).

\bibitem{schulz_moriond}
H.~J. Schulz,  in {\em Correlated fermions and transport in mesoscopic
  systems}, edited by T. Martin, G. Montambaux, and J. {Tran Thanh Van}
  (Editions fronti\`eres, Gif sur Yvette, France, 1996).

\bibitem{note_logs_1}
It is important to remark that the correlation functions have been calculated
  by using the fixed point Hamiltonian. This means that there are logarithmic
  corrections to the expressions that we quote due to asymptotic freedom. Such
  logarithmic corrections have been analyzed for instance in Refs.
  \onlinecite{affleck_log_corr,giamarchi_logs} for SU(2) spin chains.

\bibitem{note_logs_2}
The conformal transformation was carried out right at the fixed point. For a
  system that does not sit exactly at the fixed point, there would be
  corrections coming from the irrelevant operators. This is the finite
  temperature counterpart of the logarithmic corrections that are obtained at
  $T=0$.

\bibitem{golinelli_incommensurate}
O. Golinelli, T. Jolicoeur, and E. Sorensen, Incommensurability in the magnetic
  excitations of the bilinear-biquadratic spin-1 chain, 1998, cond-mat/9812296.

\bibitem{diptiman_sen_pc}
D. Sen, private communication.

\bibitem{parity}
S. Eggert and I. Affleck, Phys. Rev. B {\bf 46},  10866  (1992).

\bibitem{dmrg2}
S. Qin, J. Lou, Z. Su, and L. Yu,  in {\em Density-Matrix Renomalization}, {\em
  Lecture Note in Physics}, edited by I. Peschel, X. Wang, M. Kaulke, and K.
  Hallberg (Springer, Berlin, 1999).

\bibitem{dmrg3}
S. Qin {\it et~al.}, Phys. Rev. B {\bf 56},  9766  (1997).

\bibitem{haldane_bosons}
F.~D.~M. Haldane, Phys. Rev. Lett. {\bf 47},  1840  (1981).

\bibitem{cardy_scaling}
J.~L. Cardy, {\em Scaling and Renormalization in Statistical Physics}, {\em
  Cambridge Lecture Notes in Physics} (Cambridge University Press, Cambdridge,
  UK, 1996).

\end{thebibliography}

\begin{table}
\caption{Correspondance between the notations of the present paper and
those of K. Tandon \emph{et al.}}
\begin{tabular}{c|c|c|c|c}
 & States & & & \\
\tableline
Present paper & $\mid \tilde{1}\rangle$ & $ \mid\tilde{2}\rangle$ & $
\mid\tilde{3}\rangle $ & \\
K. Tandon \emph{et al.} & $j \mid 7'\rangle$ & $  j^2 \mid 5' \rangle$ & $ \mid 1
\rangle
$ & \\
\tableline
 & Operators & & & \\
\tableline
Present paper & $ T_1^+$ & $ T_2^+$ & $ T_3^+$ & $ T^z+1/3 $ \\
K. Tandon \emph{et al.} & $ j^2 \tau^-$ & $ j L^- $ & $j^2 R^- $ & $ \sigma^z $\\
\tableline
\end{tabular}
\label{correspondance1}
\end{table}

\begin{figure}
\centerline{\epsfig{file=fig2.eps,angle=0}}
\caption{The energy levels of a single triangle as a function of the
magnetic field. Solid lines correspond to states with S=3/2, dashed
lines to states with S=1/2. One observes the level crossing between
the state with $S^z=3/2$ and the states with $S=1/2,S^z=1/2$ as the
magnetic field is increased.}
\label{fig:levels}
\end{figure}

\begin{figure}
\centerline{\epsfig{file=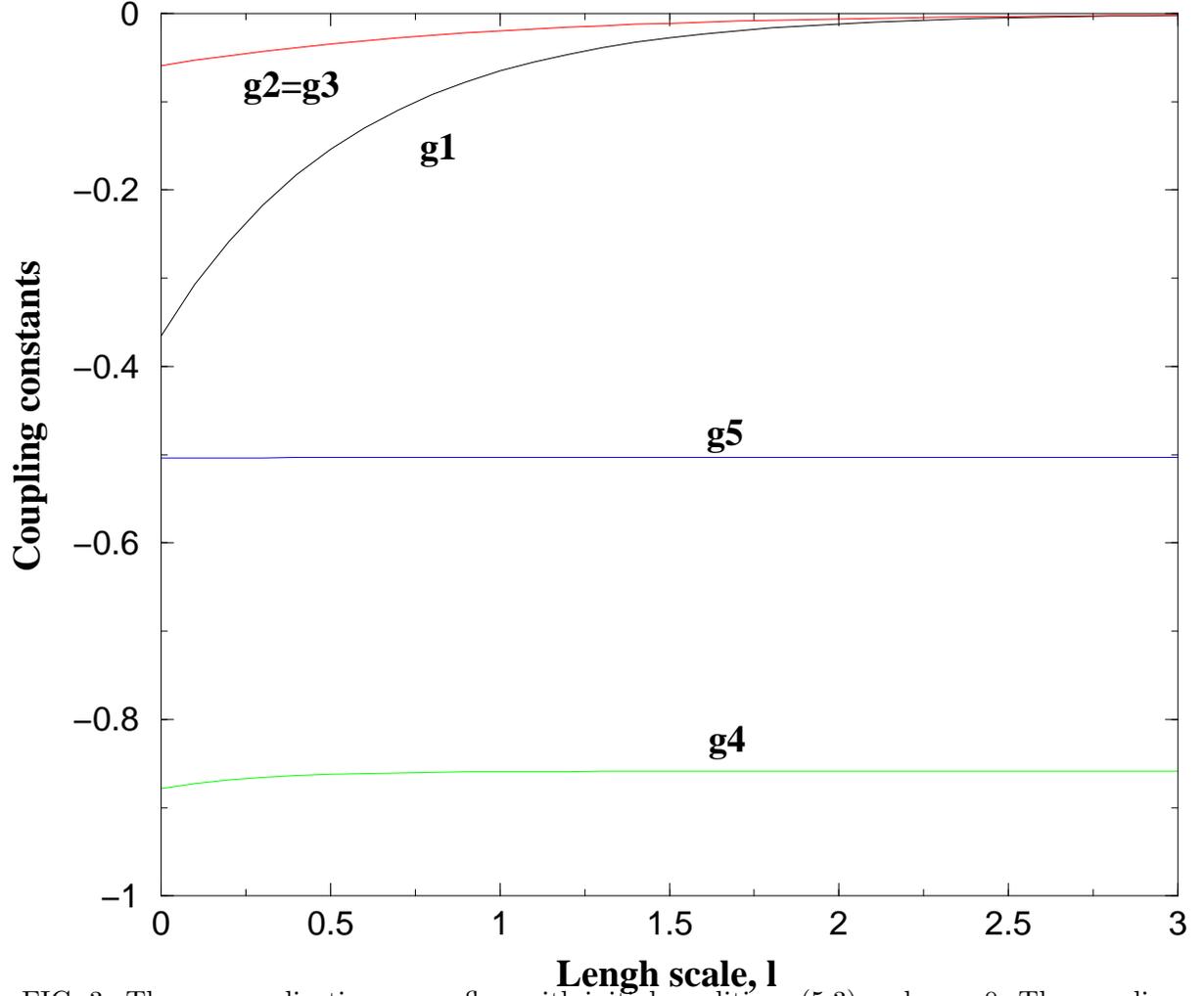,angle=0}}
\caption{The renormalization group flow with initial conditions
(\ref{eq:ini_cond}) and $y_0=0$. The couplings constants $g_1,g_2,g_3$
are renormalized to 0, whereas $g_4 \to g_4^*$, and $g_5 \to g_5^*$.
The system therefore flows to a two component Luttinger liquid fixed
point.}
\label{fig:rgflow_u=0} 
\end{figure}

\begin{figure}
\centerline{\epsfig{file=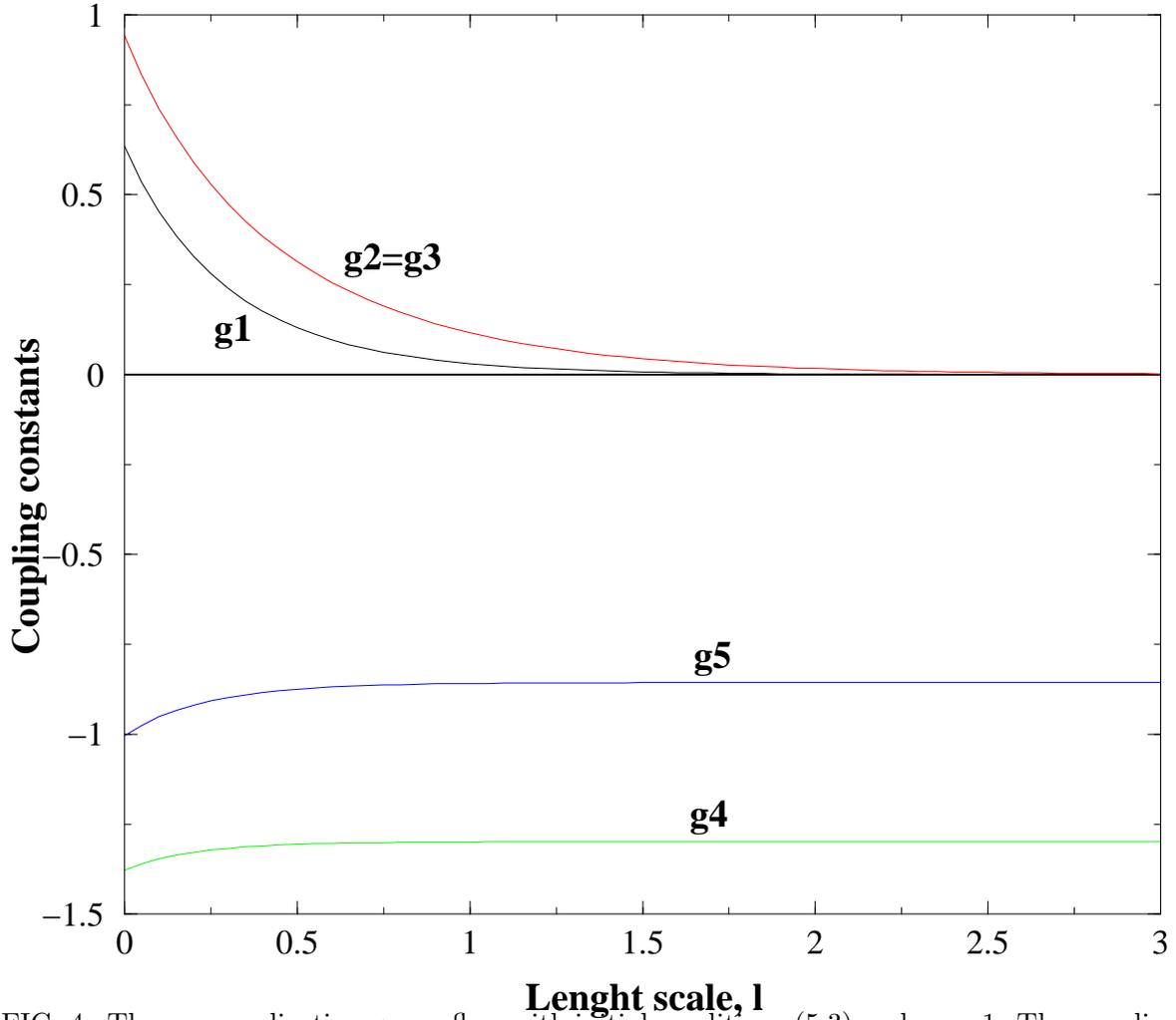,angle=0}}
\caption{The renormalization group flow with initial conditions
(\ref{eq:ini_cond}) and $y_0=1$. The couplings constants $g_1,g_2,g_3$
are renormalized to 0, whereas $g_4 \to g_4^*$, and $g_5 \to g_5^*$.
The presence of a marginal perturbation preserving SU(3) symmetry does
not suppress the two component Luttinger liquid behavior. However, by
comparing with Fig.~\ref{fig:rgflow_u=0}, it is seen that it changes
the exponents at the fixed point.}
\label{fig:rgflow_u=1} 
\end{figure}

\begin{figure}
\centerline{\epsfig{file=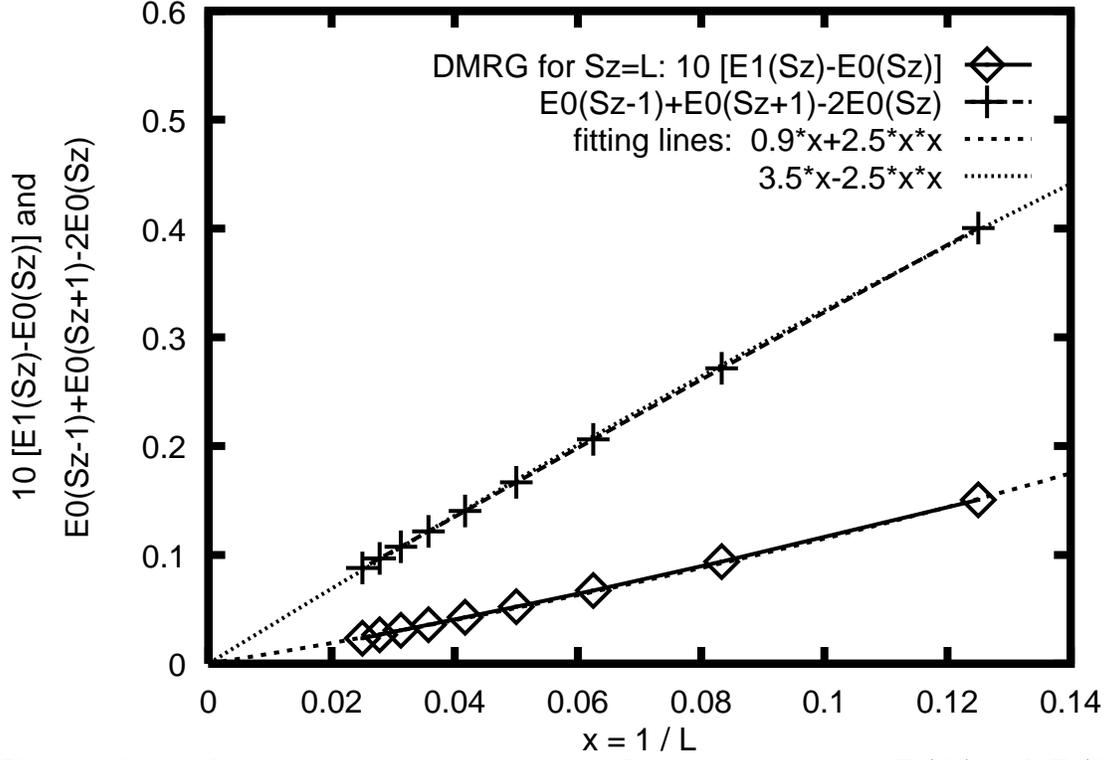,angle=0}}
\caption{
Gap scalings for magnetization $m=1$.  Lowest two energies $E_0(S_z)$ 
and $E_1(S_z)$ for each $S_z$ obtained by DMRG are plotted into 
$\Delta S^{tot,z}=0$ gap scaling: $10[E_1(S_z)-E_1(S_z)]$ vis $1/L)$, and
$\Delta S^{tot,z}=1$ gap scaling: $E_0(S_z-1)+E_0(S_z+1)-2 E_0(S_z)$ vis 
$1/L$ with $S_z=L$.  We have magnified the $\Delta S^{tot,z}=0$ gap by 
times to make the figure clear.  The linear fittings are 
$E_1(S_z)-E_0(S_z)\sim 1/L$ and 
$E_0(S_z-1)+E_0(S_z+1)-2 E_0(S_z) \sim 1/L$ in thermodynamic limit.  
The chain lengths calculated by DMRG are $L=8$, $12$, $\ldots$, $40$.
}
\label{fig:gapscaling} 
\end{figure}

\end{document}